\documentclass[12pt]{iopart}
\usepackage{graphicx}
\begin{document}
\paper[Phase lapses in quantum dots]{Transmission phase lapses in quantum 
dots: the role of dot-lead coupling asymmetry }
\author{D I Golosov$^1$ and Yuval Gefen$^2$}
\address{
 $^1$ Racah Institute of Physics,
the Hebrew University, Jerusalem 91904, Israel}
\ead{golosov@phys.huji.ac.il}
\address{$^2$ Dept. of Condensed Matter Physics, Weizmann Institute of Science,
Rehovot 76100, Israel}
\ead{yuval.gefen@weizmann.ac.il}
\begin{abstract}
Lapses of transmission phase in transport through quantum dots are
ubiquitous already in the absence of interaction, in which case
their precise location is determined by the signs and magnitudes of
the tunnelling matrix elements. However, actual measurements  for  a
quantum dot embedded in an Aharonov-Bohm interferometer  show
systematic  sequences of phase lapses  separated by  Coulomb
peaks -- an issue that attracted much attention and generated controversy. 
Using  a two-level  quantum dot as an example we show that
this phenomenon can be accounted for by the combined effect of
asymmetric  dot-lead couplings (left lead/right lead asymmetry as well as
different level broadening for different levels) and interaction-induced 
"population
switching" of the levels, rendering this behaviour generic.
We construct and analyse a mean field scheme for an interacting 
quantum dot, and investigate the properties of the mean field solution,
paying special attention to the character of its dependence (continuous
vs. discontinuous) on the chemical potential or gate voltage.
\end{abstract}
\submitto{\NJP}
\maketitle

\section{Introduction}
\label{sec:intro}

Recent systematic studies \cite{Yacoby,Schuster,Avinun} of current transmission through quantum dot (QD) 
embedded
into an arm of an Aharonov--Bohm 
interferometer \cite{Ora,GIA,YGreview,Aharony}, 
uncovered an unusual,
correlated behaviour of transmission phase as a function of the gate
voltage. Namely, between any two consecutive Coulomb 
blockade peaks
the transmission phase suffers one abrupt change (phase lapse) of $-\pi$. 

This surprising feature cannot be understood within the framework of a
non-interacting QD model \cite{BGEW,OG97,SOG,Kim}, where the 
presence or absence of a phase 
lapse
between the two transmission peaks is determined by the relative signs
of the tunnelling matrix elements coupling the corresponding QD levels to
the leads. Roughly speaking, two adjacent peaks are separated by a phase 
lapse as long as the product of the four matrix elements, coupling each
of the two levels to the two leads, is positive \cite{OG97,SOG,Kim}.
Since experimentally there is no way to control these signs in a typical 
QD \cite{Leturcq},
this would suggest an approximately 50\% probability of the presence of a 
phase lapse between the two consecutive peaks, in disagreement with
experimental data. This dictates that the Coulomb (charging) interaction
between electrons in the QD must be accounted for at some 
level \cite{YGreview,Weidenmueller,But99,Baltin,Hackenbroich}.  

Earlier investigations \cite{Baltin,SI} of the interaction effects in QDs 
resulted in the
notion of ``population switching'' between the broad (strongly coupled to
the leads) and narrow dot levels with varying gate voltage (or equivalently, 
with varying chemical potential). This phenomenon, which subsequently 
attracted attention
of both theorists \cite{BvOG,Dagotto,Marquardt,Meden,vonDelft,Kim06,Lee06,Avi} 
and 
experimentalists (see, {\it e.g.,} reference \cite{ExpInt}), 
consists in the narrow levels being
shifted upward due to the Coulomb potential of electrons accumulated at a
broad level, which remains near the Fermi energy (``hovers'') over an extended
range of the gate voltage/chemical potential values. Within the latter range, 
successive narrow levels from time to time get rapidly filled with 
electrons (thereby emptying the broad level, hence the term ``population
switching'' \cite{Baltin,Hackenbroich,SI,BvOG}) and shift downward below the 
Fermi level. Available 
results \cite{SI,YG04,Sindel} 
suggest that this switching may be either continuous or discontinuous,
although no systematic study of the two scenarios has been performed.
We note that already in references \cite{YG04,Sindel} mean field approach
(self-consistent Hartree approximation) has been employed.
Apart from reference \cite{SI}, where  a QD
with only one lead was considered, these early studies were all concerned 
with  models where the absolute
values of the tunnelling coupling of each individual level to the right
and left lead were the same (``left-right symmetry'').

It is probably due to the latter circumstance that for a long
time no attention has been paid to another generic interaction-induced 
mechanism which affects both the energy level structure and the transmission
phase behaviour in an interacting QD. This mechanism, which is not
effective in the opposite-sign left-right 
symmetric-coupling case {\it only}, in the
context of more conventional solid state physics corresponds to forming
excitonic correlations between different bands \cite{Khomskii}. Within the
mean-field approach to interacting QDs \cite{prb06}, a similar scenario 
consists
in forming the off-diagonal average values between the QD states, leading
to interaction-induced hybridisation between the dot energy levels. We note
that the importance of considering the generic, right-left asymmetric dot-lead
coupling values was pointed out independently in reference \cite{Marquardt}.

For the case of a strongly-interacting QD in the Coulomb blockade
regime one finds that even a small right-left asymmetry typically results
in a large effective hybridisation (again in analogy with conventional
impurity problems \cite{Khomskii}), leading in turn to a change
of the coupling sign for the effective dot levels 
(``minus'' changed to ``plus''), and to a presence of
phase lapse between the two transmission peaks. These findings, reported
earlier in reference \cite{prb06}, are in line with recent functional 
renormalisation group results obtained for both two-level 
\cite{Marquardt,Meden,vonDelft} and multi-level \cite{Meden} interacting dots. 
We also note the recent perturbative treatment \cite{Kim06} and
treatments based on the renormalisation group/Bethe {\it ansatz} approach \cite{Lee06,Avi}.

In the present article, we investigate transmission through a QD in the 
spinless
case at zero temperature. We begin by describing the behaviour of transmission
phase for multi-level non-interacting dot in section \ref{sec:nonint}.
The mean-field approach to an interacting two-level QD is formulated in 
section \ref{sec:model}. The remainder of the paper is devoted to
illustrating and discussing {\it the two mechanisms, whose interplay brings 
about the abrupt changes of the transmission phase (as a function of the gate 
voltage
or chemical potential) between the Coulomb blockade peaks}. First, in
section \ref{sec:exciton} we discuss the ``excitonic''
restructuring of the QD spectrum, operational in the presence of a 
right-left asymmetry. Second, in section  \ref{sec:switching} we address 
the two
scenarios (continuous and discontinuous) of ``population switching''
in the right-left symmetric case. The 
interplay between the two mechanisms is briefly described in section
\ref{sec:phases}. The overall conclusions, along with some remarks on
the generality of our results and a summary of outstanding questions, are 
relegated to section \ref{sec:conclu}.

\section{Phase Lapses in the Non-Interacting Case}
\label{sec:nonint}

{\it Transmission phase lapses in a multilevel dot in the absence of
electron-electron interaction. Dependence on the dot-lead coupling
strength. The role of dot-lead coupling signs. General expression for the 
transmission amplitude. Example: a four-level dot.}
\\

In the present section, we discuss the behaviour of transmission phase in the
case of a generic multi-level QD, in the absence of electron-electron
interactions \cite{OG97,SOG,Kim,Weidenmueller,But99}. While our analysis here 
is far from being comprehensive, it allows
to draw three important conclusions:

\noindent (i) Phase lapses represent a generic property of transmission through
QDs. In other words, the transmission amplitude $t_{tr}$  vanishing
at  certain values of the chemical potential or the gate voltage (at which 
point \cite{SOG,But99} the value of the transmission phase jumps by $-\pi$) 
does not impose
any restrictive condition on the QD parameters. We quote reference \cite{prb06}
for an analysis of what is often perceived as a contradiction between this 
statement and the Friedel sum rule. 

\noindent (ii) As mentioned in the Introduction, in the non-interacting case
the location of phase lapses with respect to transmission peaks is crucially 
affected by the relative signs of tunnelling matrix elements coupling the QD
levels to the right and left leads \cite{SOG,OG97}. In particular, a necessary 
(but possibly
not always sufficient)  condition for a single phase lapse to occur
between the two successive dot levels (corresponding to the two successive
transmission peaks) is that the product of the tunnelling
matrix elements coupling the two corresponding levels to the two leads is
negative (we stress that this can be severely modified in the presence of
electron-electron interactions and an asymmetric QD-lead coupling, see
sections \ref{sec:exciton} and \ref{sec:switching} below).

\noindent (iii) Beyond the above conclusion, the actual distribution of
transmission zeroes with respect to the dot levels strongly depends on 
the values of the QD parameters, and this dependence can be rather complex.

Other analyses of the transmission phase in the absence of interaction
\cite{BGEW} have demonstrated that disorder or geometry alone would not give rise to sequences of correlated inter-resonance phase lapses. The latter may
however result from a presence of a very broad dot level \cite{Oreg}.

\begin{figure}
\includegraphics{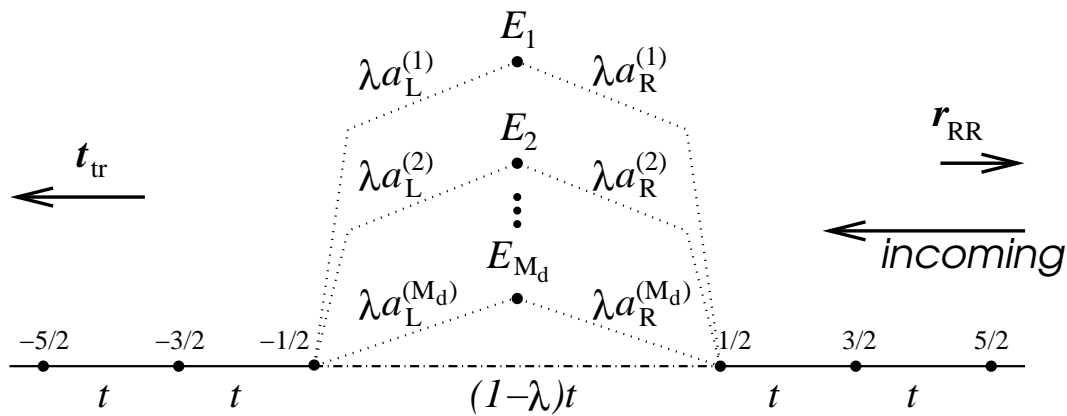}
\caption{\label{fig:schemedot} The model system, composed of a
wire (chain) and a two-level dot, equations (\ref{eq:wire}--\ref{eq:gendot}).
Fully coupled dot corresponds to $\lambda=1$. The arrows show incoming and
outgoing waves, cf. equation (\ref{eq:waves})}
\end{figure}

The Hamiltonian of a generic non-interacting dot
with $M_d$ levels $E_i$ (see figure \ref{fig:schemedot}) is given by 
\begin{equation}
{\cal H}_d=\sum_{i=1}^{M_d} \left( E_i - \mu \right) 
\hat{d}^\dagger_j \hat{d}_j 
\,.
\end{equation}
The dot is coupled to a one-dimensional wire
\begin{equation}
{\cal H}_w= -\frac{t}{2} \sum_j \left(\hat{c}^\dagger_j
\hat{c}_{j+1} + \hat{c}^\dagger_{j+1} \hat{c}_j \right)- \mu
\sum_j \hat{c}^\dagger_j \hat{c}_j, \label{eq:wire}
\label{eq:Hwire}
\end{equation}
(where $t$ is half bandwidth and the half-integer index $j$ labels the sites
in the wire), 
by the perturbation $V$,
\begin{eqnarray}
V&=&V_T+\frac{t}{2}\left(\hat{c}^\dagger_{1/2}\hat{c}_{-1/2}+
\hat{c}^\dagger_{-1/2}\hat{c}_{1/2}  \right), 
\label{eq:tunnel2} \\
V_T&=&-\frac{1}{2}\sum_{i=1}^{M_d}\hat{d}^\dagger_j \left(a_{L}^{(i)} \hat{c}
_{-1/2} +
a_R^{(i)} c_{1/2} \right)+ {\rm H.c.}\,,
\end{eqnarray}
where $a^{(i)}_{L}$ ($a^{(i)}_{R}$) are the real tunnelling elements coupling
the $i$th dot level to the left (right) lead; these are assumed to be
small in comparison to the bandwidth in the wire, $2t$.  The second term on 
the r. h. s.
in equation (\ref{eq:tunnel2})
corresponds to cutting the link between  sites $j=-1/2$ and
$j=1/2$ of the wire. It is instructive to consider the system with varying
perturbation strength, $\lambda$,
\begin{equation}
{\cal H}_\lambda={\cal H}_d+{\cal H}_w+\lambda V
\label{eq:gendot}
\end{equation}
(figure \ref{fig:schemedot}).
While for $\lambda =1$ this corresponds to a fully coupled dot (no direct 
hopping between right and left leads), characterised by a certain sequence
of transmission peaks and transmission phase lapses, at $\lambda \rightarrow 0$
one recovers a featureless transmission amplitude, $t_{tr}(\mu) \equiv 1$
(hence $\Theta_{tr} \equiv 0$),
of the decoupled case 
(unperturbed wire). As $\lambda$ increases from 0 to 1, the profile of 
 $\Theta_{tr}(\mu)$ evolves  between these two limiting cases.

When searching for an eigenfunction $\psi$ which away from the dot
has the form of a left-moving, partially reflected wave,
\begin{equation}
\psi(x)=\left\{ \begin{array}{ll}t_{tr}\exp (-{\rm i}kx),\,\,& x \leq -1/2,\\ & \\
\exp ({\rm -i}kx) + r_{RR} \exp ({\rm i}kx), \,\,\,&x \geq 1/2,
\end{array} \right. 
\label{eq:waves}
\end{equation}
(where $t_{tr}$ and $r_{RR}$ are transmission and right-right reflection
amplitudes respectively)
one has to solve a system of $M_d+2$ linear equations for the quantities
$1/t_{tr}$, $\,r_{RR}/t_{tr}$, and the 
values $\psi_i$, $i = 1,...,M_d$ of the wave function on the dot levels
(these correspond to the Schr\"{o}dinger equation ${\cal H}_\lambda \psi =
-t\psi\cos k$ written for the wire sites $j=\pm 1/2$ and for the dot levels).
Zeroes of the transmission coefficient $t_{tr}$ (at $\lambda \neq 0$) 
obviously coincide
with those of the principal determinant $D$ of this linear system.
It is straightforward to find that
\begin{equation} 
D \propto \sin k \left\{2(1-\lambda)t- \lambda^2 
\sum_{i=1}^{M_d} \frac{a_L^{(i)} a_R^{(i)}}{\epsilon-E_j} \right\}
\prod_{l=1}^{M_d}(\epsilon-E_l)\,,
\label{eq:determinant}
\end{equation} 
where $\epsilon=-t\cos k$. We will now analyse this expression in the two
limiting cases.

\noindent $\lambda \ll 1$ -- {\it weakly coupled dot}. As $\lambda$ decreases, 
the zeroes $\epsilon=Z_i$ approach the dot levels, $\epsilon = E_i$. To leading
order in $\lambda$, we find $Z_i= E_i+\lambda^2 a_L^{(i)} a_R^{(i)}/(2t)$. Thus,
(i) the direction in which the transmission zero is shifted with respect to
the corresponding dot level $E_i$ is determined solely by the sign of 
the product of the tunnelling elements, $\sigma_i={\rm sign} a^{(i)}_L 
a^{(i)}_R$.
We thus find that whenever all such signs are the same, each inter-level
energy interval has exactly one transmission zero, whereas otherwise there
arise intervals which contain either two such zeroes or none at all.
(ii) Not counting the spurious zero at $k=0$, there are exactly $M_d$ 
transmission zeroes. At $\lambda\rightarrow 0$, each zero approaches the
corresponding level $E_i$, cancelling its transmission peak and thereby
restoring the featureless $t_{tr}(\mu)$ [and $\Theta_{tr}(\mu)$] at 
$\lambda =0$.

\noindent $\lambda \rightarrow 1$ -- {\it fully coupled dot}. As the value of $\lambda$ 
increases,
the transmission zeroes move further away from their corresponding dot levels.
While at a finite $\lambda$ they obviously cannot cross any of the dot
levels to drift into  neighbouring inter-level intervals, in those intervals 
which 
contain pairs of zeroes the zeroes may meet and disappear from the real energy
axis; likewise, new pairs of zeroes may appear in some inter-level intervals. 
In addition, one or more zeroes which at smaller $\lambda$ may have been 
located
below the lowest dot level or above the highest one may now move out of the 
conduction band. 
In the fully coupled case of $\lambda=1$,
the r.\ h.\ s. of equation (\ref{eq:determinant}) 
is proportional to the polynomial of the power 
$M_d-1$, which guarantees that the maximal possible number of transmission
zeroes is $M_d-1$; this number increases to $M_d$ for $0< \lambda < 1$ [it
follows that at $\lambda \rightarrow 1$ a transmission zero located outside
the dot energy range must move out of the conduction band along the real
energy axis]. 
When the two adjacent levels are characterised 
by
the same signs $\sigma_i=\sigma_{i+1}$, it is certain that there is a
transmission zero  (or possibly an odd number of zeroes) between them; 
otherwise there may be either two zeroes 
(or, in principle, an even number of zeroes) or none.

It appears that the possibility to have more than two zeroes in any inter-level
interval requires a fine-tuning of the parameters of QD, and cannot be 
considered as generic. Barring these special cases, we see that 
had we compared this unrealistic non-interacting scenario to the
experimentally observed correlated occurrence of exactly one transmission
zero between every two adjacent dot levels, this would indeed require all the
level coupling signs, $\sigma_i$, to be the same.

Furthermore, when at $\lambda=1$ for one of the dot levels $E_i$ the quantity 
$|a^{(i)}_L a^{(i)}_R|$ is much smaller than the energy distances to the 
neighbouring levels, such a weakly coupled level is approached by a 
single transmission zero at
\begin{equation}
Z_i=E_i- a^{(i)}_L a^{(i)}_R \left[\sum_{l\neq i}  \frac{a^{(l)}_L a^{(l)}_R}
{E_i -E_l}
\right]^{-1}.
\end{equation}
We note that in this case, the direction from which the transmission zero
would approach is not determined solely by $\sigma_i$.

Finally, for reference purposes we quote the full expression for the 
transmission coefficient of a generic dot as given by equation 
(\ref{eq:gendot}),
\begin{eqnarray} \fl t_{tr}^{(\lambda)}(\mu)=-{\rm i} \sqrt{t^2-\mu^2} 
\left[2(1-\lambda)t 
+ \lambda^2\sum_{i=1}^{M_d}\frac{a^{(i)}_L a^{(i)}_R}{E_i-\mu} \right]
\left\{ {\rm i} (2\lambda -\lambda^2-2)t 
\sqrt{t^2-\mu^2}
\begin{array}{c}\mbox{}\\ \mbox{} \\ \mbox{} \end{array} \right.-
\nonumber \\
-\lambda^2 \sum_{i=1}^{M_d}\frac{(a^{(i)}_L)^2 t+(a^{(i)}_R)^2 t- 2 
(1-\lambda)
(\mu-{\rm i} \sqrt{t^2-\mu^2})a^{(i)}_L a^{(i)}_R}{2(E_i-\mu)} +
\nonumber \\ \left.
+\mu(\lambda^2-2\lambda)t-\lambda^4\frac{\mu-{\rm i}\sqrt{t^2-\mu^2}}{4t} \sum_{i<l} \frac{(a^{(i)}_R 
a^{(l)}_R
-a^{(l)}_R a^{(i)}_L)^2}{(E_i-\mu)(E_l-\mu)}\right\}^{-1}\,.
\label{eq:dotreference}
\end{eqnarray}
Equation (\ref{eq:dotreference}) contains the full information about the
locations of phase lapses and transmission peaks. The latter are shifted
with respect to the dot energy levels $E_i$, but the values of these
shifts contain a pre-factor of the order of $|a_L^{(j)}a_R^{(l)}|/t^2 \ll 1$
and in most cases of interest can be treated as small. 

In particular, we find that at $\lambda=1$, and  
$|a^{(i)}_{L,R}| \ll |E_i-E_l|$
for all $i$ and $l$, the transmission phase $\Theta_{tr}(\mu)$ in the 
vicinity of a dot level $E_i$ is given by
\begin{equation}
\tan \Theta_{tr}(\mu) = \frac{\mu}{\sqrt{t^2-\mu^2}}+ \frac{(a^{(i)}_L)^2+(a^{(i)}_R)^2}
{2 \sqrt{t^2-\mu^2}
\left[E_i-\mu-\sum_{l\neq i} \frac{\left(a^{(i)}_R a^{(l)}_L
-a^{(l)}_R a^{(i)}_L\right)^2}{4(E_l-E_i)t^2}\right]}\,.
\label{eq:closeddot}
\end{equation}
This corresponds to the expected smooth increase of the transmission phase
by $\pi$ with the value of $\mu$ increasing and sweeping  $E_i$. When the
chemical potential lies away from the band edges, $|\mu|<1$, the width of 
the step is of the order of the level broadening\footnote{This is given
by $4\Gamma_i=\pi [(a^{(i)}_L)^2+(a^{(i)}_R)^2]\rho(1/2)$, where $\rho(1/2)$
is the local density of states at the terminal point of a lead, {\it e.g.,} 
$d\langle\hat{c}^\dagger_{1/2} \hat{c}_{1/2}\rangle/d \mu$ at $\lambda=1$.},
\begin{equation}
\Gamma_i=\frac{(a^{(i)}_L)^2+(a^{(i)}_R)^2}{2t^2}\sqrt{t^2-\mu^2}\,,
\label{eq:broadening}
\end{equation}
and the sum in the denominator of equation (\ref{eq:closeddot}) shifts the 
position of the centre of this
step (coinciding with the transmission amplitude maximum) on the energy axis 
with respect to the bare dot level $E_i$.

\begin{figure}
\includegraphics{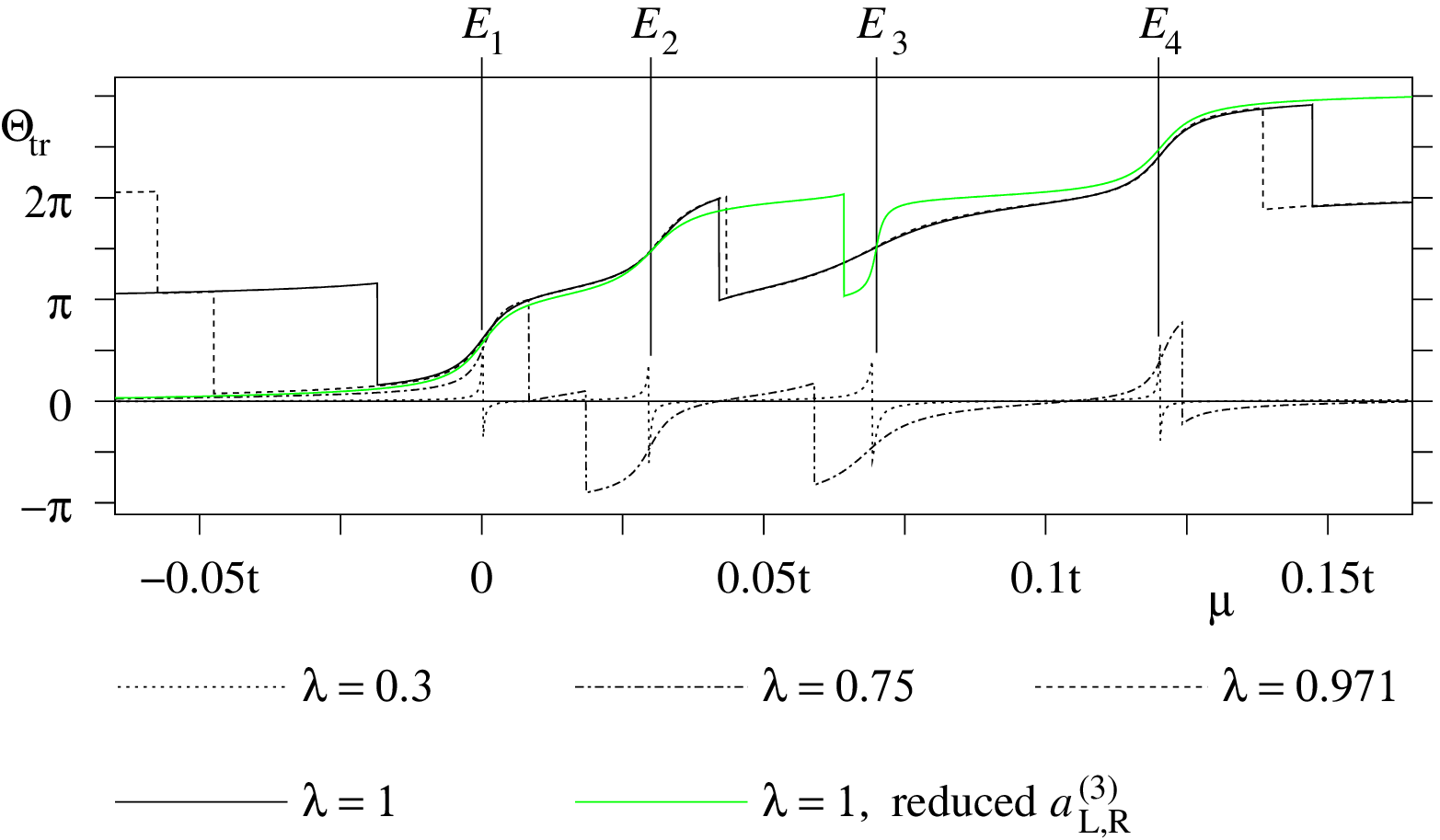}
\caption{\label{fig:phasenoU} (colour)
The complex behaviour of the transmission phase for different regimes of
QD-lead coupling.
The dependence of $\Theta_{tr}$ on $\mu$ for a non-interacting four-level 
dot with $a^{(i)}_L/t=\{0.05, 0.07, 0.1, 0.08\}$ and $a^{(i)}_R/t=\{0.08, 
-0.08, -0.12, 0.06\}$. The dot spectrum
is given by $E_i/t=\{0, 0.03, 0.07, .12\}$. The ``perturbation strength'' 
$\lambda$ (cf. figure \ref{fig:schemedot}) is equal to $0.3$, $0.75$ $0.971$, 
and $1$
for dotted, dashed-dotted, dashed, and solid black lines respectively.
The green solid line corresponds to the fully coupled dot ($\lambda=1$) with
reduced values of the tunnelling matrix elements for the 3rd level: 
$a_L^{(3)}=0.03 t$ and $a_R^{(3)}= -0.036t$.}
\end{figure}

This complex behaviour of transmission phase is exemplified by 
figure \ref{fig:phasenoU} for the case of a four-level dot with $\sigma_i=
\{1,-1,-1,1\}$. At a relatively small value of $\lambda =0.3$ (dotted line),
transmission phase remains close to zero except in the immediate vicinity
of the dot levels, and there is a phase lapse of $\pi$ near every level.
With increasing $\lambda$, the continuous increase of $\Theta_{tr}$ by $\pi$
in the vicinity of each level becomes progressively less steep, and the 
phase lapses shift further away from the levels. The directions and rates
of these shifts  reflect the differences in the coupling signs $\sigma_i$ and 
coupling magnitudes. While for all values of $\lambda$ there is exactly
one phase lapse between $E_2$ and $E_3$ and none between $E_3$ and $E_4$,
the pair of phase lapses located between $E_1$ and $E_2$ approach each other,
as shown by the dashed-dotted line ($\lambda=0.75$). They eventually merge and
disappear, and with further increase of $\lambda$ a new pair of transmission
zeroes emerges outside the dot energy range at $\mu < E_1$ (see the dashed
line, $\lambda=0.971$). With increasing $\lambda$, one of these new phase 
lapses moves towards $E_1$, whereas the other moves rapidly to the left,
disappearing in the fully coupled case of $\lambda=1$ (solid black line). 
The solid green line illustrates the effect of reducing the coupling of
a single dot level $E_3$ to the leads in the $\lambda=1$ case. We see that
the increase of transmission phase, $\Theta_{tr}(\mu)$, 
by $\pi$ near $E_3$ becomes steeper as we reduce $a^{(3)}_{R,L}$, and
a phase lapse approaches $E_3$ from the left, ``annihilating'' the smooth
increase in the limit $a^{(3)}_{R,L} \rightarrow 0$. We also note that with
decreasing coupling to the third QD level, the two of phase lapses located to 
the left
of $E_1$ and to the right of $E_4$ move away from the dot levels and disappear.

In the general case of a non-interacting multi-level dot, the behaviour of 
$\Theta_{tr}(\mu,\lambda)$ reflects the analytical properties of the complex
transmission amplitude $t_{tr}$, which are probably not known in detail.
In any case, these properties are far too cumbersome to try and analyse the 
effect of charging interaction on $\Theta_{tr}$ in such a generic case. One is
therefore left with the option to consider the effects of interaction in 
a simple model case in order to identify the relevant 
correlation-induced {\it mechanisms} with the hope that the results will prove
generic at a qualitative level.  

\section{The Mean Field Scheme for an Interacting Two-Level Dot}
\label{sec:model}

{\it The model for a two-level interacting QD. Rotation of the fermionic
operators on the dot and the ``intra-dot hopping'' as a measure of right-left
asymmetry. Mean field decoupling and mean field equations. Invariance of
the mean field scheme with respect to the choice of the basis. Properties
of the effective non-interacting system.}
\\

In this section, we describe our method of calculation of the transmission
phase, $\Theta_{tr}(\mu)$, for an interacting two-level QD. After introducing
the model, we focus on the mean field decoupling and on the properties of
the resultant non-interacting system.

The minimal model for studying the phase lapse mechanism includes
a two-level QD,
\begin{equation}
{\cal H}_{QD} = (E_1^{(0)}-\mu ) \hat{d}^\dagger_1 \hat{d}_1 + (
E_2^{(0)}-\mu ) \hat{d}^\dagger_2 \hat{d}_2+ U \hat{d}^\dagger_1
\hat{d}^\dagger_2 \hat{d}_2 \hat{d}_1\,.
\label{eq:Hdotgen}
\end{equation}
Here, the operators $\hat{d}_i$ with $i=1,2$ annihilate  electrons
on the two dot sites (with bare energies $\{E^{(0)}_i\}$,
$E^{(0)}_2>E^{(0)}_1$). The QD is coupled to the two leads by the
tunnelling term
\begin{equation} \fl
V_T=- \frac{1}{2} \hat{d}_1^\dagger \left( a_L \hat{c}_{-1/2}+
a_R \hat{c}_{1/2} \right)-\frac{1}{2} \hat{d}_2^\dagger 
\left(b_L \hat{c}_{-1/2}+b_R
\hat{c}_{1/2} \right)+ {\rm h. c.}.
\label{eq:tunnel}
\end{equation}
The operators $\hat{c}_j$ (with half-integer $j$) are defined on
the tight-binding sites of the left and right lead (cf. figure
\ref{fig:schemedot}). In the $U=0$ case, the location of the transmission
phase lapse with respect to the dot levels is determined by the coupling
sign, $\sigma^{(0)}={\rm sign} \,a_R a_L b_R b_L$; in particular, in the 
opposite-sign case of $\sigma^{(0)}=-1$ the phase lapse occurs outside the
inter-level energy interval.

In spite of the simplicity of this model, (\ref{eq:Hdotgen}--\ref{eq:tunnel}),
no exact analytical solution for general values of parameters has been
found so far. In a recent paper \cite{prb06}, the present writers suggested a 
mean field approach to this problem. Here, we will further explore
the generality of our mean field scheme and the physical properties
of the mean field solutions.

\begin{figure}
\includegraphics{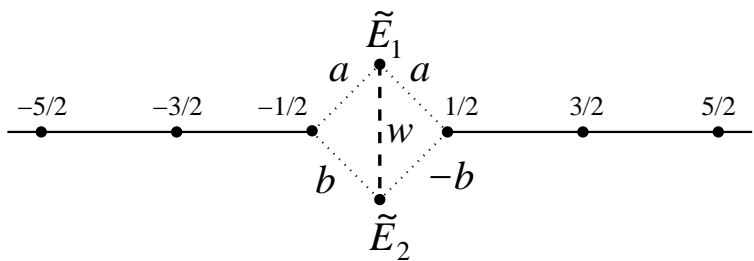}
\caption{\label{fig:schemedot2l} The model system, composed of a
wire (chain) and a two-level dot, equations (\ref{eq:tunnel1}) and
(\ref{eq:intradot}).}
\end{figure}

At the outset, we confine ourselves to studying the case when the values of
tunnelling couplings in equation (\ref{eq:tunnel}) obey the
following condition:
\begin{equation}
b_R^2-b_L^2=a_L^2-a_R^2\,,
\label{eq:constraint}
\end{equation}
which corresponds to a certain 3D subspace of the full space of all values
of $a_{L,R}$ and $b_{L,R}$ (the latter are assumed to be real). While we 
did not attempt an investigation of the case when the constraint 
(\ref{eq:constraint}) is not satisfied, there is an expectation that no
significant physics is missed by making this assumption (possibly barring
a few isolated singular cases), which however offers an important technical
benefit. Indeed, by making an orthogonal transformation of the QD fermion
operators,
\begin{equation}
\left(\begin{array}{ll} \hat{d}_1 \\ \hat{d}_2 \end{array} \right) =
O \left(\begin{array}{ll} \tilde{d}_1 \\ \tilde{d}_2 \end{array} \right)\,,
\,\,\,\,\,\,O=\left(\begin{array}{ll} \cos \varphi & -\sin\varphi \\ 
\sin \varphi & \cos \varphi \end{array} \right)
\label{eq:orthogonal}
\end{equation}
with $\tan \varphi =(a_R-a_L)/(b_L-b_R)$ we find  that the tunnelling term,
\begin{eqnarray} 
V_T=&-&\frac{1}{2} \left\{(a_L \hat{c}_{-1/2}^\dagger+ 
a_R \hat{c}^\dagger_{1/2}) (\cos \varphi \tilde{d}_1 -\sin \varphi \tilde{d}_2)
+ \right. \nonumber \\
&+& \left. (b_L \hat{c}_{-1/2}^\dagger+ 
b_R \hat{c}^\dagger_{1/2}) (\sin \varphi \tilde{d}_1 +\cos \varphi \tilde{d}_2)
\right\} + {\rm H. c.}\,\,,
\nonumber
\end{eqnarray}
then reduces to a simple form,
\begin{equation} \fl
V_T=- \frac{1}{2} a \left( \hat{c}^\dagger_{-1/2}+
\hat{c}^\dagger_{1/2}\right) \tilde{d}_1-\frac{1}{2} b  
\left( \hat{c}^\dagger_{-1/2}-
\hat{c}^\dagger_{1/2} \right)\tilde{d_2}+ {\rm H. c.}.
\label{eq:tunnel1}
\end{equation}  
At the same time, the transformation (\ref{eq:orthogonal}) changes the form
of the dot Hamiltonian, equation (\ref{eq:Hdotgen}), to
\begin{equation} \fl
{\cal H}_{QD}=(\tilde{E}_1^{(0)}-\mu ) \tilde{d}^\dagger_1 \tilde{d}_1 +
(\tilde{E}_2^{(0)}-\mu ) \tilde{d}^\dagger_2 \tilde{d}_2-
\frac{w_0}{2} ( \tilde{d}^\dagger_1 \tilde{d}_2 + \tilde{d}^\dagger_2
\tilde{d}_1 )+ U \tilde{d}^\dagger_1
\tilde{d}^\dagger_2 \tilde{d}_2 \tilde{d}_1\,.
\label{eq:intradot}
\end{equation}
Thus, the system can be formally viewed as a  quantum dot with the ``site
energies'' $\tilde{E}_{1,2}^{(0)}$ and the ``intra-dot hopping'' $w_0$;
coupling to the leads is now left-right symmetric, with the ``coupling
sign'' $\tilde{\sigma} =-1$ (figure \ref{fig:schemedot2l}). 
Further 
analysis will be carried out in terms of this new QD, assuming
$\tilde{E}^{(0)}_2>\tilde{E}^{(0)}_1$. We note that
the quantity 
\begin{equation}
w_0=\frac{1}{4}\, \frac{(E_2^{(0)}-E_1^{(0)}) (b_L-b_R)(a_L-a_R)}
{a_L (a_L-a_R)+b_L (b_L-b_R)}
\end{equation}
is in reality a measure of left-right asymmetry of the original dot coupling,
equation (\ref{eq:tunnel}); by varying $w_0$, one can probe both opposite 
sign ($\sigma^{(0)}=-1$) and same-sign ($\sigma^{(0)}=1$) cases (see section 
\ref{sec:exciton} below).

The mean-field calculation  entails decoupling the interaction term in
equation (\ref{eq:intradot}),
\begin{eqnarray}
\tilde{d}^\dagger_1 \tilde{d}^\dagger_2 \tilde{d}_2 \tilde{d}_1
&\!\rightarrow &\!\tilde{d}^\dagger_1 \tilde{d}_1 \langle
\tilde{d}^\dagger_2 \tilde{d}_2 \rangle + \tilde{d}^\dagger_2
\tilde{d}_2 \langle
\tilde{d}^\dagger_1 \tilde{d}_1 \rangle - \langle \tilde{d}^\dagger_1
\tilde{d}_1 \rangle \langle
\tilde{d}^\dagger_2 \tilde{d}_2 \rangle- \nonumber \\
&&\!-\tilde{d}^\dagger_1 \tilde{d}_2 \langle \tilde{d}^\dagger_2
\tilde{d}_1 \rangle-
\tilde{d}^\dagger_2 \tilde{d}_1 \langle \tilde{d}^\dagger_1
\tilde{d}_2 \rangle +
|\langle \tilde{d}^\dagger_1 \tilde{d}_2 \rangle|^2\,,
\label{eq:HFdecomp}
\end{eqnarray}
which, when substituted in equation (\ref{eq:intradot}), yields an effective
non-interacting dot,
\begin{equation}
{\cal H}_d^{MF} = (\tilde{E}_1-\mu) \tilde{d}^\dagger_1 \tilde{d}_1 +
(\tilde{E}_2-\mu)
\tilde{d}^\dagger_2 \tilde{d}_2
-\frac{w}{2}
(\tilde{d}^\dagger_1 \tilde{d}_2 + \tilde{d}^\dagger_2 \tilde{d}_1).
\label{eq:hammf}
\end{equation}
The self-consistency conditions take the form of three coupled mean-field
equations,
\begin{eqnarray}
\tilde{E}_1 = \tilde{E}_1^{(0)} &+U& \langle \tilde{d}_2^\dagger
\tilde{d} _2 \rangle ,
\,\,\,\,\,\,
\tilde{E}_2 = \tilde{E}_2^{(0)} + U \langle \tilde{d}_1^\dagger
\tilde{d}_1 \rangle ,
\label{eq:mfee2} \\
&w =&w_0 + 2 U \langle{\tilde{d}^\dagger_1 \tilde{d}_2} \rangle.
\label{eq:mfew}
\end{eqnarray}

Here, the relevant average values are obtained from the 
thermodynamic potential, $\Omega_{MF}$, of the effective non-interacting
system with the Hamiltonian,
\begin{equation}
{\cal H}_{tot}^{MF}={\cal H}_d^{MF}+{\cal H}_{w}+V_T
+\frac{t}{2}\left(\hat{c}^\dagger_{1/2}\hat{c}_{-1/2}+
\hat{c}^\dagger_{-1/2}\hat{c}_{1/2}  \right)
\end{equation}
[cf. equations (\ref{eq:wire}--\ref{eq:tunnel2})] according to
\begin{equation}
\langle \tilde{d}^\dagger_{1} \tilde{d}_{1} \rangle =
{\partial \Omega_{MF}}/
{\partial \tilde{E}_{1}}\,, \,\,\,\, \langle \tilde{d}^\dagger_{2} 
\tilde{d}_{2} \rangle =
{\partial \Omega_{MF}}/
{\partial \tilde{E}_{2}}\,, \,\,\,\, \langle \tilde{d}^\dagger_1 \tilde{d}_2
\rangle = -{\partial \Omega_{MF}}/{\partial w} \,.
\label{eq:averages}
\end{equation}
An exact and convenient method to evaluate $\Omega_{MF}$ is described
in reference \cite{prb06}.

Clearly, the advantage of the choice (\ref{eq:constraint}) lies in the fact
that at the mean-field level, interaction effects amount to a rather 
simple renormalisation of the coefficients in the single-particle part
of the Hamiltonian, equation (\ref{eq:intradot}). We note that a similar 
mean field scheme
can be constructed for a symmetric same-sign QD, when the constraint
(\ref{eq:constraint}) is replaced by $a_L=a_R$, $b_L=b_R$. This case,
which is analysed in reference \cite{prb06}, will not be discussed here;
we note also that a unitary transformation of the dot operators allows
one to recast the corresponding Hamiltonian in the form 
(\ref{eq:tunnel1}--\ref{eq:intradot}) with $b=0$.

It is important to notice that had we carried out the decoupling, equation 
(\ref{eq:HFdecomp}),
in any other basis of the dot operators (including the original one,
that of $\hat{d}_{1,2}$), the resultant system of mean field equations
would have been equivalent to (\ref{eq:mfee2}--\ref{eq:mfew}).
This means that the two operations: rotating the basis [cf. equation
(\ref{eq:orthogonal})] and performing the Hartree -- Fock decoupling
[equation (\ref{eq:HFdecomp})] are commutative. 
Indeed, consider an arbitrary orthogonal transformation of the dot
operators,
\begin{equation}
\left(\begin{array}{ll} \tilde{d}_1 \\ \tilde{d}_2 \end{array} \right) =
O^\prime \left(\begin{array}{ll} {d}^\prime_1 \\ {d}^\prime_2 \end{array} 
\right)\,,
\,\,\,\,\,\,O^\prime=\left(\begin{array}{ll} \cos \phi & -\sin\phi \\ 
\sin \phi & \cos \phi \end{array} \right),
\label{eq:orthogonal2}
\end{equation}
in the QD Hamiltonian, equation (\ref{eq:intradot}).
Then in the new basis of operators $d_{1,2}^\prime$ we have:
\begin{equation}
\tilde{E}_1^{(0)\prime}=\tilde{E}_1^{(0)} \cos^2 \phi + \tilde{E}_2^{(0)} 
\sin^2 \phi - \frac{1}{2}w_0 \sin 2\phi\,,
\end{equation}
and similar equations for $\tilde{E}_1^{(0)\prime}$ and $w_0^\prime$;
the interaction term remains unchanged. 
Now suppose that the ``renormalised'' quantities $\tilde{E}_{1,2}$ and $w$
satisfy the mean field equations, (\ref{eq:mfee2}),(\ref{eq:mfew}), and 
(\ref{eq:averages}), and do the same transformation $O^\prime$ {\it after}
the decoupling (\ref{eq:HFdecomp}), {\it i.e.,} in the
mean field Hamiltonian (\ref{eq:hammf}), yielding
\begin{eqnarray}
\tilde{E}_1^\prime&=&\tilde{E}_1 \cos^2 \phi + \tilde{E}_2 
\sin^2 \phi - \frac{1}{2}w \sin 2\phi\,,
\label{eq:e1prime}\\
\tilde{E}_2^\prime&=&\tilde{E}_1 \sin^2 \phi + \tilde{E}_2 
\cos^2 \phi + \frac{1}{2}w \sin 2\phi\,,\\
w^\prime&=&(\tilde{E}_1-\tilde{E}_2) \sin 2 \phi + w \cos 2 \phi\,.
\label{eq:wprime}
\end{eqnarray}   
On the other hand, had we chosen to perform the decoupling in the 
$d^\prime_{1,2}$ basis, we would have obtained the mean field equations
\begin{eqnarray}
\tilde{E}_1^\prime = \tilde{E}_1^{(0)\prime} &+U& {\partial \Omega_{MF}}/
{\partial \tilde{E}_{2}^\prime},
\,\,\,\,\,\,
\tilde{E}_2^\prime = \tilde{E}_2^{(0)\prime} + U {\partial \Omega_{MF}}/
{\partial \tilde{E}_{1}^\prime} ,
\label{eq:mfee2p} \\
&w^\prime =&w_0^\prime - 2 U {\partial \Omega_{MF}}/{\partial w^\prime}.
\label{eq:mfewp}
\end{eqnarray}
It remains to verify that the quantities $E_{1,2}^\prime$ and $w^\prime$ as 
obtained from
equations (\ref{eq:e1prime}--\ref{eq:wprime}) solve the 
system (\ref{eq:mfee2p}-\ref{eq:mfewp}). Since
\begin{equation}
\frac{\partial \Omega_{MF}}{\partial \tilde{E}_1^\prime} = 
\frac{\partial \Omega_{MF}}{\partial \tilde{E}_1}\cos^2 \phi +
\frac{\partial \Omega_{MF}}{\partial \tilde{E}_2}\sin^2 \phi-
\frac{\partial \Omega_{MF}}{\partial w} \sin 2 \phi
\end{equation}
etc., this is easily done by a direct inspection. We conclude that the two 
systems of mean field equations, (\ref{eq:mfee2}--\ref{eq:mfew}) and
(\ref{eq:mfee2p}--\ref{eq:mfewp}) are equivalent to each other, hence the
results are indeed independent on the choice of basis. This statement 
illustrates the fact that our mean-field approximation is a conserving one \cite{Baym},
and holds also if one replaces $O^\prime$ in equation (\ref{eq:orthogonal2})
with an arbitrary unitary matrix.  

As we will see below, under certain conditions one finds that for a given
value of $\mu$, the mean field equations (\ref{eq:mfee2}--\ref{eq:mfew})
for $\tilde{E}_1$, $\tilde{E}_2$, and $w$ may have multiple solutions.
In this case, one must choose the solution which corresponds to the lowest
value of the full thermodynamic potential including the constant (non-operator)
terms in equation (\ref{eq:HFdecomp}),
\begin{equation}
\Omega=\Omega_{MF}-U\langle \tilde{d}^\dagger_1 \tilde{d}_1 \rangle \langle
\tilde{d}_2^\dagger \tilde{d}_2\rangle +U \langle \tilde{d}_1^\dagger 
\tilde{d}_2\rangle^2\,.
\label{eq:omegatot}
\end{equation}
While the invariance of $\Omega_{MF}$ under the transformation 
(\ref{eq:orthogonal2}) is obvious, it is straightforward to check that
the sum of the last two terms on the r. h. s. also does not change, so that
the entire quantity $\Omega$ is independent on the choice of basis.

Once the values of mean field parameters $E_{1,2}$ and $w$ are determined,
the transmission phase $\Theta_{tr}$ 
[corresponding to the effective non-interacting model
(\ref{eq:hammf})] can be evaluated from
\begin{equation}
\sqrt{t^2-\mu^2}\,{\rm tan} \Theta_{tr}(\mu)
=\mu+\frac{b^2(\tilde{E}_1-\mu)+
a^2(\tilde{E}_2-\epsilon)+2\mu a^2b^2/t^2}
{(\tilde{E}_1-\mu)(\tilde{E}_2-\mu)-
\frac{1}{4}w^2-a^2b^2/t^2}\,.
\label{eq:phaseint}
\end{equation} 
As a function of $\mu$, $\Theta_{tr}$ shows two smooth steps of $+\pi$ 
centred at the transmission peaks, $\mu=\mu_{1,2}$ with
\begin{equation}
\mu_{1,2}=(\tilde{E}_1+\tilde{E}_2)/2 \mp \frac{1}{2}
[(\tilde{E}_1-\tilde{E}_2)^2+w^2+4a^2b^2/t^2]^{1/2}\,.
\label{eq:intdotpeaks}
\end{equation}
More precisely, at $\mu=\mu_{1,2}$ the quantity $\Theta_{tr}/\pi$ has 
half-integer values. Friedel sum rule then implies \cite{prb06} that the same 
holds
for the electron population change due to the dot, $N_{dot}=N(\mu)-N_w(\mu)$,
where $N(\mu)$ is the total number of carriers in the system, and $N_w(\mu)$
is the number of carriers at an unperturbed (connected) wire alone
[see equation (\ref{eq:Hwire})], evaluated at the same value of $\mu$.
 
The positions of these peaks $\mu_{1,2}$ are slightly shifted outwards 
with respect to the
mean-field dot energy levels [eigenvalues of (\ref{eq:hammf})]:
\begin{equation}
E_{1,2}=(\tilde{E}_1+\tilde{E}_2)/2 \mp \frac{1}{2}
[(\tilde{E}_1-\tilde{E}_2)^2+w^2]^{1/2}\,.
\label{eq:intdoteigen} 
\end{equation}
Equation (\ref{eq:phaseint}) determines $\Theta_{tr}$ up to a shift by a 
multiple of $\pi$. One can easily investigate the evolution of 
$\Theta_{tr}(\mu)$ for the effective non-interacting model (\ref{eq:hammf})
as the ``interaction strength'' $\lambda$ [cf. equation 
(\ref{eq:gendot})] varies from 0 to 1. We thus find that this shift
should be chosen in such a way that 
\begin{equation}
\Theta_{tr} (\mu \rightarrow -t) \rightarrow \left\{ \begin{array}{ll}
-\pi/2 &  {\rm for~~} a^2>b^2,\\
-3\pi/2 & {\rm for~~} a^2<b^2. \end{array} \right.
\end{equation}
In addition, at $\mu=Z$ with
\begin{equation} 
Z= \frac{\tilde{E}_2 a^2-\tilde{E}_1 b^2}{a^2-b^2}\,.
\label{eq:intdotzero}
\end{equation}
the transmission phase suffers a lapse of $-\pi$ [{\it i.e.,} $\Theta_{tr}$
includes a term, $-\pi\,\theta(\mu-Z)$].

In the range of parameters where multiple solutions to the mean field equations
arise, it is possible that, {\it e.g.}, $\mu=Z$ or $\mu=\mu_{1,2}$ corresponds
to a thermodynamically unstable solution. This situation and its implications
for the overall $\Theta_{tr}(\mu)$ profile will be discussed in greater detail
in sections \ref{sec:switching} and \ref{sec:phases}. We shall now proceed 
with analysing the
properties of mean field solutions for various values of parameters.

\section{First Mechanism for Abrupt Phase Change between Transmission Peaks: 
Effects of Left-Right Asymmetry And Excitonic Correlations}
\label{sec:exciton}

{\it Overview of mean field results: ``phase diagram''. Excitonic mechanism
in ``phase'' 1: eigenstates of the dot are linear combinations of site
states, hence effective
coupling sign change due to off-diagonal correlations on the dot. Evolution
of the effective dot parameters with varying chemical potential $\mu$. 
Excitonic mechanism becomes ineffective when approaching either the 1-2 or
left-right symmetric situations.}
\\

\begin{figure}
\includegraphics{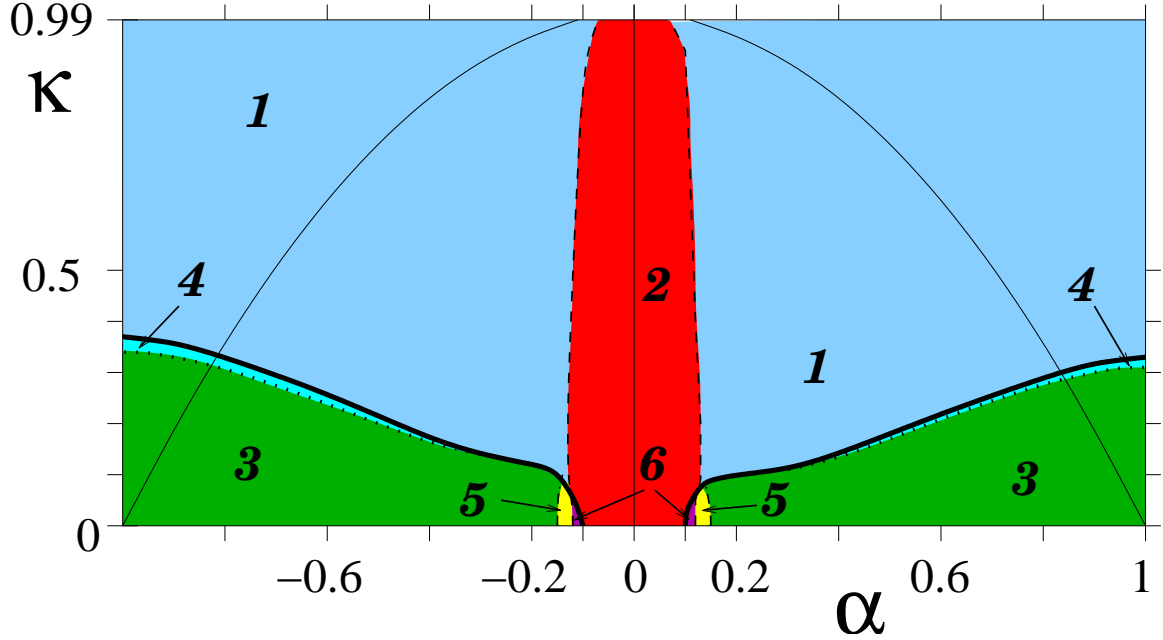}
\caption{\label{fig:phadiag} (colour)
The ``phase diagram'' of a two-level QD shown in figure \ref{fig:schemedot2l}.
The parameters are $U=0.1t$, $\tilde{E}_1^{(0)}=0$, $\tilde{E}_2^{(0)}=0.004t$,
and $\sqrt{a^2+b^2}=0.125t$.
The axes represent
the 1-2 level asymmetry, $\alpha = (|a|-|b|)/\sqrt{a^2+b^2}$,
and the dimensionless intra-dot hopping, $\kappa=w_0/
[(\tilde{E}_1^{(0)}-\tilde{E}_2^{(0)})^2+w_0^2]^{1/2}$.}
\end{figure}

We begin our discussion of the mean-field results for a 
two-level QD with presenting a typical mean-field ``phase diagram''
(figure \ref{fig:phadiag}). This shows how the location
of the transmission phase lapse  $Z$ with respect to the two transmission peaks
$\mu_{1,2}$ [see equations (\ref{eq:intdotpeaks})and (\ref{eq:intdotzero})] 
depends on the
two dimensionless QD parameters,
\begin{equation}
\alpha = \frac{|a|-|b|}{\sqrt{a^2+b^2}}\,,\,\,\,\kappa=\frac{w_0}
{\sqrt{(\tilde{E}_1^{(0)}-\tilde{E}_2^{(0)})^2+w_0^2}}\,.
\label{eq:phadiagparameters}
\end{equation} 
Of these, $\alpha$ is a measure of the 1-2 level broadening asymmetry, whereas
$\kappa$ is related to left-right asymmetry of the levels coupling to the 
two leads. In this section, we will be interested in the generic situation
when $\kappa$ is sufficiently large.  In figure \ref{fig:phadiag}, this 
corresponds to ``phases'' 1 and 2 (blue and red).  We will see that this 
typically gives rise to a hybridisation between the dot levels, which in 
turn results in the phase lapse of $-\pi$ occurring between the two 
peaks (``phase'' 1). ``Phase'' 2, which occupies a narrow stripe near 
the 1-2 symmetric case, is characterised by the occurrence of a $-\pi$-phase
lapse outside the region $\mu_1 < \mu < \mu_2$. 

The physics underlying this correlated restructuring of the QD spectrum at
larger $\kappa$ is thus somewhat similar to that of exciton formation in a 
semiconductor. A closer analogy can be drawn with the 
``excitonic correlations'' between localised and extended states in impurity
problems \cite{Khomskii}. At smaller $\kappa \rightarrow 0$ we encounter
another mechanism, that of ``population switching'' as discussed earlier
for QDs in the Coulomb blockade regime \cite{Baltin,SI,YG04,Sindel} 
(section \ref{sec:switching}). The entire ``phase diagram'', figure 
\ref{fig:phadiag}, can be interpreted in the context of interpolation
between these two regimes (section \ref{sec:phases}).

Let us first formally examine the role of left-right asymmetry in our
mean-field scheme, starting with equation (\ref{eq:HFdecomp}). 
There, the diagonal and off-diagonal terms contain two types of quantum
average values,
\begin{equation}
\tilde{n}_{1,2} = \langle\tilde{d}_{1,2}^\dagger \tilde{d}_{1,2} \rangle\,,
\,\,\,\,\tilde{n}_{12}=\langle \tilde{d}_{1}^\dagger \tilde{d}_{2} \rangle\,.
\end{equation}
In terms of the ``transformed'' QD, equation (\ref{eq:intradot}), these 
correspond to the two dot site occupancies, and to the intra-dot ``excitonic''
\cite{Khomskii} hybridisation respectively.

Let us first suppose that the value of $w_0$ is sufficiently small, so that in
the absence of interaction the coupling of the two QD eigenstates 
(\ref{eq:intdoteigen}) to the dot is opposite-sign, $\sigma^{(0)}=-1$:
\begin{equation}
\fl
w_0^{\,2}< (2Z^{(0)}-\tilde{E}^{(0)}_1-\tilde{E}_2^{(0)})^2
- (\tilde{E}^{(0)}_1-\tilde{E}_2^{(0)})^2\,,\,\,\,\,\,
Z^{(0)}= \frac{\tilde{E}_2^{(0)} a^2-\tilde{E}_1^{(0)} b^2}{a^2-b^2}\,.
\label{eq:woppsign}
\end{equation}
In terms of figure \ref{fig:phadiag}, this corresponds to the area below
the thin solid line. We recall that in this situation, the non-interacting
dot would have the phase lapse located outside the inter-level energy
interval [although strictly speaking it may still barely fall  between 
the two transmission peaks, which are slightly shifted with respect to the
dot levels, equations (\ref{eq:intdoteigen}) and (\ref{eq:intdotpeaks})]. 

As will be discussed in more detail below, the typical situation in the
large-U case is that, due to large values of $\tilde{n}_{12}$ in equation
(\ref{eq:mfew}),  $w$ is strongly renormalised in the relevant energy region 
around the phase lapse. If the value of $w$ becomes sufficiently large, the 
effective dot sign $\sigma$ will change from $-1$ to $1$, for this is
obviously the case for 
\begin{equation}
|w| \gg |\tilde{E}_2 -\tilde{E}_1|
\label{eq:largew}
\end{equation} 
[when the product of the couplings of the two effective dot eigenstates, 
$(\tilde{d}_1 \pm \tilde{d}_2)/\sqrt{2}$, to the leads is given by 
$(a^2-b^2)^2/4>0$]. 
The phase lapse will then be located between the two mean-field dot levels, 
$E_{1}< Z < E_2$.

\begin{figure}
\includegraphics{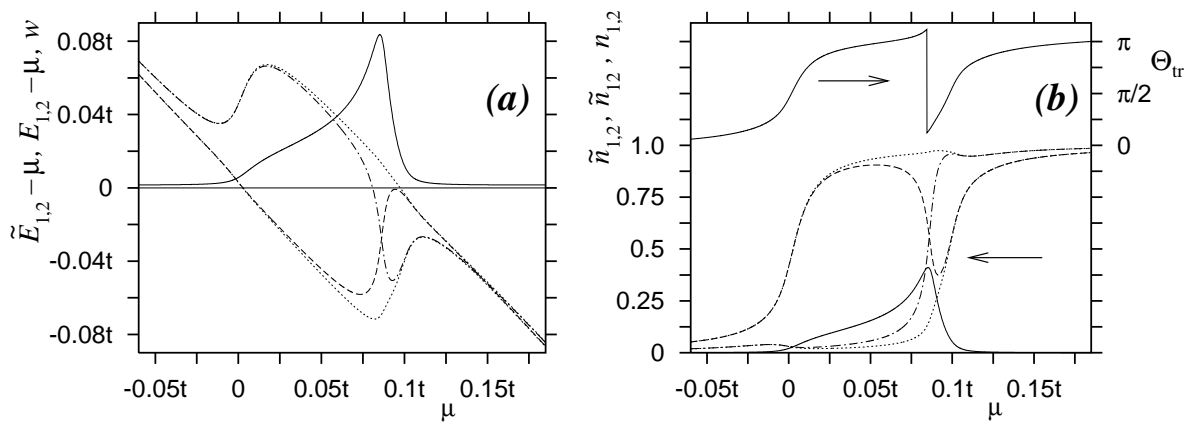}
\caption{\label{fig:exciton} Mean field properties of a QD with $\alpha=0.3$,
$\kappa=0.35$,  $U=0.1t$, $\tilde{E}_1^{(0)}=0$, $\tilde{E}_2^{(0)}=0.004t$,
and $\sqrt{a^2+b^2}=0.125t$ plotted vs. chemical potential $\mu$. 
{\it (a)} ``Intra-dot hopping'' $w$ (solid line) and the differences between 
mean field  ``site energies'' and chemical potential, 
$\tilde{E}_{1,2}-\mu$ (dashed and dashed-dotted lines). Dotted lines show the 
differences between the mean field energy levels and chemical potential, 
$E_{1,2}-\mu$. {\it (b)} ``Off-diagonal average'' $\tilde{n}_{12}$ (lower 
solid line), ``site occupancies'' $\tilde{n}_{1,2}$ (dashed and dashed-dotted 
lines), and the mean-field level occupancies $n_{1,2}$ (dotted lines). The 
upper solid line shows the transmission phase, $\Theta_{tr}$ (right scale). }
\end{figure}

This ``excitonic'' mechanism of the QD sign change is operational within 
the blue region
of our ``phase diagram'' (figure \ref{fig:phadiag}, ``phase'' 1), which is 
defined as the parameter region where the phase lapse of $-\pi$ occurs between 
the two 
transmission peaks and the properties of the effective dot vary continuously
with $\mu$. We see that this blue region extends well below the line
denoting the change of the original sign, $\sigma^{(0)}$, which means
that this situation is rather generic.  It is further exemplified by
figure \ref{fig:exciton}, showing the evolution of the dot properties with 
$\mu$ for a particular choice of parameters.

When the chemical potential lies well below the dot levels, the latter
are unoccupied and the dot parameters are close to their bare values.
In particular, the ``intra-dot hopping'' $w$ [solid line in figure
\ref{fig:exciton} {\it (a)}]  is close to $w_0=0.0015t$, which is small
in comparison with the difference between the two ``site energies'' 
$\tilde{E}_{1,2}$ (dashed and dashed-dotted lines), $\tilde{E}_{2}-\tilde{E}_1
\approx \tilde{E}_{2}^{(0)}-\tilde{E}_1^{(0)}=0.004t$. This ensures that
coupling to the leads is opposite-sign, $\sigma=\sigma^{(0)}=-1$ [cf. equation
(\ref{eq:woppsign})]. The wave function of
the lower QD eigenstate in this regime is mostly localised on the site
1 of the QD, which for our choice of $\alpha=0.3>0$ is the one that is 
stronger coupled to the leads ($a>b$, see figure \ref{fig:schemedot2l});
the lower level is therefore broader than the upper one [equation 
(\ref{eq:broadening})].

With increasing $\mu$, the population of this level [and hence the occupancy
of site 1, dashed line in  figure \ref{fig:exciton} {\it (b)}] begins to
grow. The Coulomb repulsion term, $U\tilde{n}_1$ in the second 
equation (\ref{eq:mfee2}) pushes the energy of the other site, $\tilde{E_2}$, 
upwards, and its population $\tilde{n}_2$ [dashed-dotted line in  
figure \ref{fig:exciton} {\it (b)}] remains low. The energy $\tilde{E}_1$
(which is very close to the lower QD energy level, $E_1$) eventually crosses 
the chemical potential, resulting in a smooth increase of $\tilde{n}_1$ and of
the transmission phase
$\Theta_{tr}$  [upper solid line in figure \ref{fig:exciton} {\it (b)}]
in agreement with the Friedel sum rule \cite{prb06}.

While the average of the two dot energy levels, $(E_1+E_2)/2$
[see equation (\ref{eq:intdoteigen})] does not depend on $w$, the energy
of the lower level $E_1$ goes down when $w$ increases, making such an
increase energetically favourable in the partially-occupied regime
of $0<\tilde{n}_1+\tilde{n}_2 <2$. We see that indeed the value of $w$ begins
to increase with increasing $\mu$, which is accompanied by  an increase of 
hybridisation [off-diagonal average value $\tilde{n}_{12}$, lower solid line 
in 
figure \ref{fig:exciton} {\it (b)}]. The nature of eigenstates begins to
change continuously, and it is no longer possible to identify the lower
eigenstate with site 1; at the same time, a large difference arises between
the site energies $\tilde{E}_{1,2}$ and the energy eigenvalues $E_{1,2}$ [dotted
lines in figure \ref{fig:exciton} {\it (a)}]. While both the site energies
$\tilde{E}_{1,2}$ and site occupancies $\tilde{n}_{1,2}$ eventually 
cross\footnote{We find $\tilde{n}_1=\tilde{n}_2$ at $\mu \approx  0.0859 t$.
The hybridisation reaches a maximum of $\tilde{n}_{12} \approx 0.411$ at
$\mu \approx  0.0852 t$, where $\tilde{n}_1-\tilde{n}_2\approx 0.1$.},
which might look reminiscent of the usual population switching 
scenario \cite{Baltin,SI,YG04,Sindel} (see also section \ref{sec:switching}), 
the eigenvalues $E_{1,2}$ never
coincide with each other, and the same holds for the respective level 
occupancies,
\begin{equation}
{n}_{1,2} = \frac{1}{2}(\tilde{n}_1 + \tilde{n}_2) \pm \frac{1}{2}
\sqrt{(\tilde{n}_1 - \tilde{n}_2)^2+4(\tilde{n}_{12})^2}
\label{eq:intdotoccupancies}
\end{equation}
[dotted lines in figure \ref{fig:exciton} {\it (b)}].

In particular, near the crossing point $\tilde{E}_1=\tilde{E}_2$ the level 
energies $E_{1,2}$ are pushed far apart by a large $w$, which peaks in this
region, thereby satisfying condition (\ref{eq:largew}). Hence the dot becomes
same-sign, $\sigma = 1$, and when (also in this region of values of $\mu$) the 
$-\pi$-phase lapse occurs with chemical potential crossing the transmission
zero, the latter is located between the two level energies/transmission peaks.

With a further increase of $\mu$, the population of the QD increases towards
$\tilde{n}_1+\tilde{n}_2=2$, and the energy gain associated with large $w$
becomes less pronounced. The value of $w$ thus begins to decrease towards
$w_0$, and it is in this region that the second level crossing, $E_2=\mu$,
occurs, accompanied by another smooth increase in $\Theta_{tr}$. The site
energies $\tilde{E}_{1,2}$  eventually cross again at $\mu \approx 0.145 t$, 
reverting to their original order.  

The presence of the $\sigma^{(0)}=1$ area in figure \ref{fig:phadiag} (above
the thin solid line) is indicative of the fact that our Hamiltonian, equation
(\ref{eq:intradot}), which is characterised by the opposite-sign ``site 
coupling'', $\tilde{\sigma}=-1$, is suitable for probing both same-sign and
opposite-sign original level coupling cases. We note that in  
figure \ref{fig:phadiag}, most of the $\sigma^{(0)}=1$ area is occupied by 
``phase'' 1.

The mechanism of interaction-induced coupling sign change becomes ineffective 
when 
approaching the line $\alpha=0$, corresponding to equal absolute values
of coupling of the two QD sites to the leads. 
The reasons for this are two-fold:
(i) a stronger increase in $w/|\tilde{E}_2-\tilde{E}_1|$ is required to reach
the coupling sign change in this case.
(ii) in parallel with the usual ``population switching'' scenario, when the 
difference in the broadening of the two levels becomes smaller, larger
$U$ is needed to make the values of $\tilde{E}_1$ and $\tilde{E}_2$ cross.
Thus if $|\alpha|$  is decreased while $U$ is kept constant, the site energies
$\tilde{E}_{1,2}$ cease to cross, and in the partially-filled QD regime
stay progressively further away from each other. This decreases the level
overlap and hence the ability of the system to form $\tilde{n}_{12}$ and 
thereby renormalise $w$. At the same time, the energy gain associated with
larger $w$ in the partially-filled region becomes smaller.

Thus the coupling sign change does not occur for a relatively narrow ``red''
region (``phase'' 2) around the $\alpha=0$ line in figure \ref{fig:phadiag}, 
where the phase lapse of $-\pi$ is located outside the energy
interval between the two transmission peaks, $\mu_1<\mu< \mu_2$. ``Phase'' 2
is also characterised by a continuous evolution of the dot properties with
varying $\mu$. The area occupied by ``phase''2 becomes smaller with increasing
$U$ or $|w_0|$; it is located below the line denoting the sign change $\sigma^{(0)}=-1 \rightarrow \sigma^{(0)}=1$.

This ``excitonic'' mechanism also breaks down when approaching the $w_0=0$
line. The reason for this is clear from figure \ref{fig:schemedot2l}: at
$w=0$, the contribution of the two virtual hopping paths (via the lead 
sites $\pm 1/2$) cancel each other, owing to the difference in the signs of 
coupling of site 2 to the right and left leads. Thus if we start with the
$w_0=0$ case, a non-zero off-diagonal average value $\tilde{n}_{12}$ cannot
be formed, and $w$ remains equal to zero for all values of $\mu$ [see 
equation (\ref{eq:mfew})]. Equivalently, in the $w_0=0$ case the dot sites
1 and 2 are coupled to even and odd combinations of electron wave functions in
the two leads respectively; phases of these combinations remain fully 
independent of each other.
 
It is, however, precisely in this region of small $w_0$ that the ``population
switching'' mechanism becomes operational in its conventional form. In order
to further understand the structure of our ``phase diagram'', 
figure \ref{fig:phadiag}, we now have to proceed with a more detailed analysis
of the $w_0=0$ case.

\section{Second Mechanism for Abrupt Phase Change between Transmission Peaks:
Population Switching in the Symmetric Case -- 
Discontinuous vs. Continuous Scenario}
\label{sec:switching}

{\it ``Population switching'' in the right-left symmetric case. Evolution
of the effective dot parameters in the discontinuous case. 
Multiple mean field solutions and the phase lapse renormalisation. 
Origins of the discontinuity. Continuous population switching in the
absence of the Coulomb blockade. Effects of population switching and excitonic
correlations on the transmission amplitude. Summary: excitonic correlations,
continuous and discontinuous population switching.}
\\

We will now consider the behaviour of an interacting two-level dot in the 
right-left symmetric case of $w_0=0$. This situation was addressed earlier 
\cite{Baltin,SI,YG04,Sindel}, and the associated notions of ``hovering level'' 
\cite{Baltin} or ``population switching'' \cite{SI} were advanced in the
literature. Nevertheless until recently \cite{prb06,Goldstein} 
no clear distinction has been made 
between the two scenarios of continuous and discontinuous population switching.
We will see that the difference between these two behaviours affects the
magnitude of transmission phase lapse. More generally, the behaviour of
a right-left symmetric QD turns out to be qualitatively different from the one
found in the larger $w_0$ regime (coupling sign change due to ``excitonic'' 
correlations, section \ref{sec:exciton}). Once we clarify the effects
of interaction at $w_0=0$, the entire ``phase diagram'', figure 
\ref{fig:phadiag}, can be understood in terms of interpolation between these
two regimes (section \ref{sec:phases}).   

The mean-field analysis reported here  has its obvious shortcomings. It is 
known (\cite{Marquardt,Meden,vonDelft,Kim06,Lee06,Avi}, some of these 
references include approximate methods) that at  
least within a certain parameter
range the discontinuity is smoothened (see also section \ref{sec:conclu} 
below).  We include our mean-field results here as a convenient reference 
point for more elaborate
analyses. We also note that  available studies of the effects of quantum
fluctuations \cite{Marquardt,Meden,vonDelft,Kim06,Lee06,Avi}  are not 
sufficient to conclude that the discontinuous
evolution as described below is {\it always} an artifact of mean-field, to 
be "cured" in a more  proper treatment.

We already pointed out that in the $w_0=0$ case the off-diagonal average
value vanishes identically, $\tilde{n}_{1,2}=0$, resulting in turn in the
absence of effective intra-dot hopping, $w=0$ [equation (\ref{eq:mfew})]. The 
``site energies''
therefore coincide with the mean-field QD energy levels, $E_{1,2}=
\tilde{E}_{1,2}$, and the same holds for site and level occupancies,
$\tilde{n}_{1,2}=n_{1,2}$ [see equations (\ref{eq:intdoteigen}) and
(\ref{eq:intdotoccupancies})]. Since there is no hybridisation, the
eigenstates of the QD do not change for all values of $\mu$, and their
respective couplings to the leads remain constant (no interaction-induced
sign change can occur). Thus, the coupling of our QD to the leads remains 
opposite-sign, so that the transmission
zero $Z$, equation (\ref{eq:intdotzero}), always lies outside the energy 
interval between the mean field QD
energy levels, on the side of the weaker-coupled level ({\it i.e.}, for
$|a|<|b|$ we find $Z<E_{1,2}$ if $E_1<E_2$, and $Z>E_{1,2}$ in the opposite
case of $E_1>E_2$).

\begin{figure}
\includegraphics{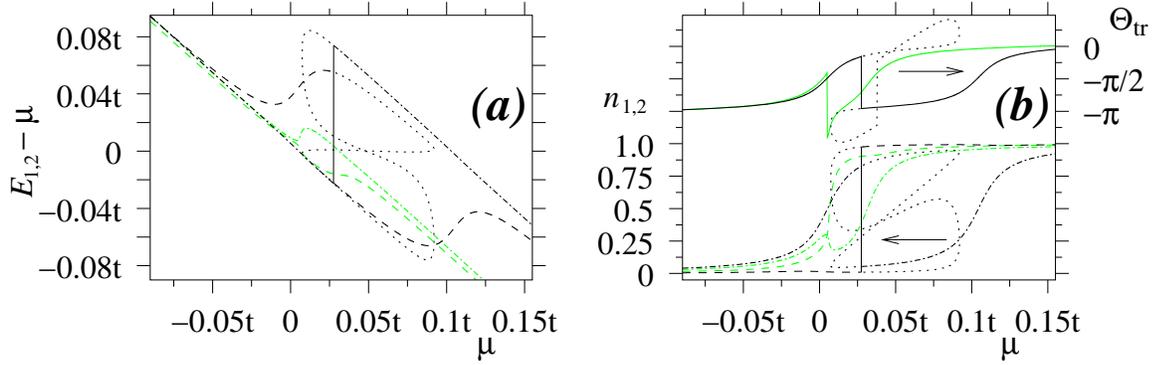}
\caption{\label{fig:switching} (colour) Mean-field properties of a 
two-level QD with right-left symmetry, plotted vs. chemical potential $\mu$ 
(black lines).  
QD parameter values are $\alpha=-0.6$, $w_0=0$,  $U=0.1t$, 
$\tilde{E}_1^{(0)}=0$, $\tilde{E}_2^{(0)}=0.004t$, and $\sqrt{a^2+b^2}=0.125t$.
 {\it (a)} the differences between 
mean field  energy levels and chemical potential, 
$E_{1,2}-\mu$ (dashed and dashed-dotted lines). {\it (b)} Level occupancies,
$\tilde{n}_{1,2}$ (dashed and dashed-dotted lines). The solid line shows 
the transmission phase, $\Theta_{tr}$ (right scale). In both {\it (a)} and
{\it (b)}, black dotted and vertical solid lines denote the unstable
solution and the discontinuous change of the QD state respectively, whereas
the respective green lines correspond to the continuous-evolution 
case of smaller $U$ and smaller $|\alpha|$, {\it viz.} $U=0.03t$, 
$\alpha = -0.25$. }
\end{figure}

\begin{figure}
\includegraphics{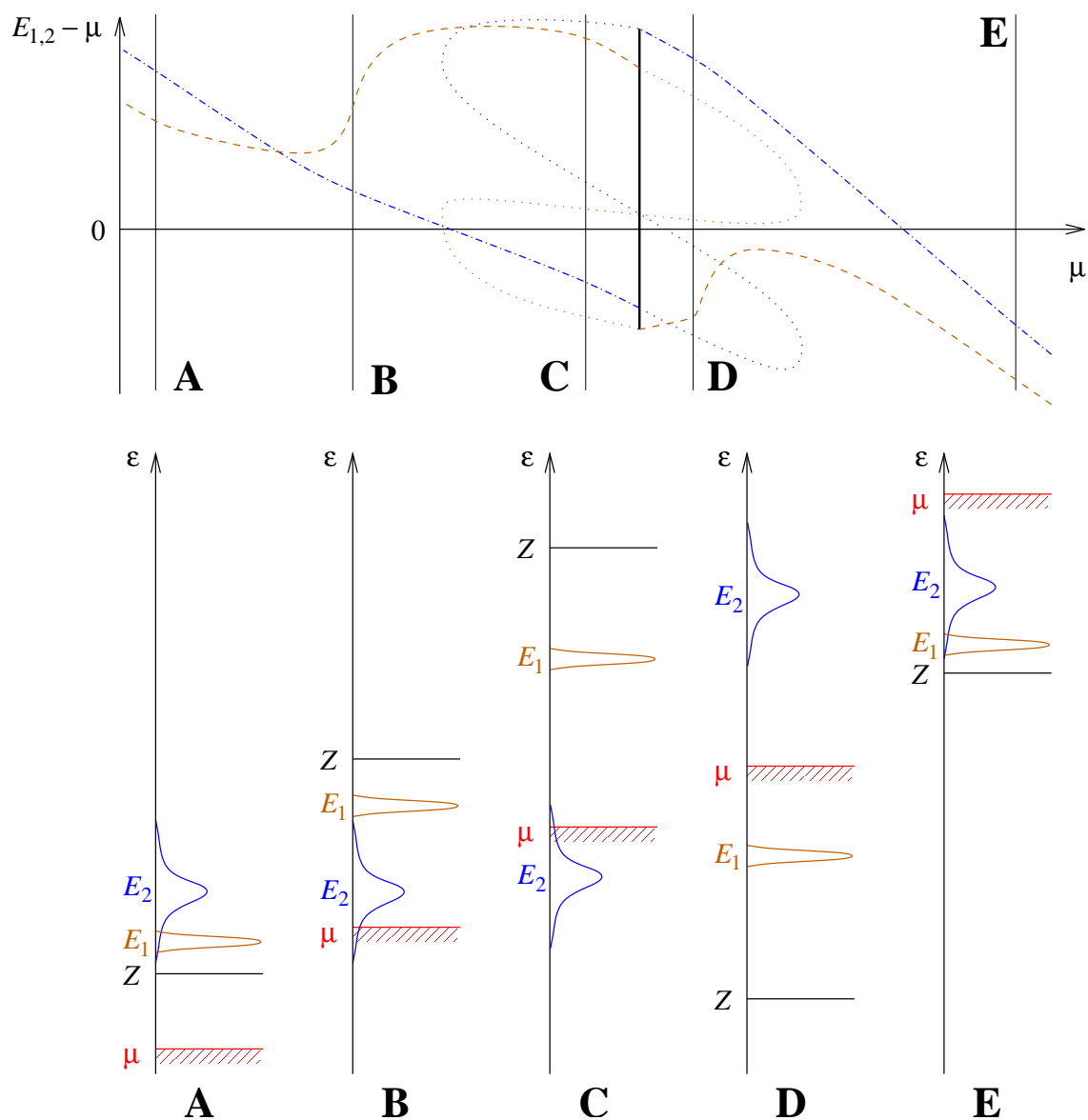}
\caption{\label{fig:switchcartoon} (colour) 
Discontinuous population switching: schematic representation of the chemical
potential dependence of the two level energies $E_{1,2}$ (dashed and 
dashed-dotted lines; it is assumed that the dot-lead couplings of levels 2 and 
1 satisfy $b>a$). Dotted lines correspond to
the unstable solution, whereas the discontinuity is marked by a bold vertical 
line. In the lower part, locations of $E_{1,2}$, $\mu$ and the transmission
zero $Z$ are depicted schematically for five points A-E, showing explicitly 
the QD level rearrangement along with the switching in the location of $Z$.}
\end{figure}

Typical behaviour of $E_{1,2}(\mu)$ and 
$n_{1,2}(\mu)$, as well as that of the transmission phase, $\Theta_{tr}(\mu)$,
is shown in figure \ref{fig:switching}. The figure corresponds to the 
$\alpha<0$ case, when the coupling of the upper bare dot level (dot site 2) 
to the leads is stronger than that of the lower one, $|b|>|a|$ 
[equation (\ref{eq:phadiagparameters})]. We will now trace the changes of
mean-field QD parameters with increasing $\mu$, addressing first the case
of stronger interaction effects (larger values of $U$ and $|\alpha|$), 
as shown in figure \ref{fig:switching} by the black lines.  
We also refer to figure \ref{fig:switchcartoon} for a schematic representation
of the corresponding changes in the mutual orientations of the two dot 
levels $E_{1,2}$, transmission zero $Z$ and chemical potential $\mu$.

\noindent (i) When the chemical potential is well below the bare QD energy
levels, $\mu < E^{(0)}_{1,2}$, the dot contribution to the overall density
of states at the Fermi level comes mainly from the tail of the broader level
$E_2$. Therefore the occupancy of the dot site 2 [black dashed-dotted line in
figure \ref{fig:switching} {\it (b)}] has a larger value, and with increasing
$\mu$ increases at a higher rate, than that of site 1 (black dashed line).
Owing to the large value of $U$, the mean field equations (\ref{eq:mfee2})
then dictate that the mean field QD level energies $E_1$ and $E_2$ [dashed
and dashed-dotted lines in figure \ref{fig:switching} {\it (a)}] eventually
cross, $E_1>E_2$ for $\mu> - 0.8 t$. The smooth increase of the dot population,
and hence of the transmission phase $\Theta_{tr} (\mu)$ [black solid line
in figure \ref{fig:switching} {\it (b)}], continues beyond this crossing point.
(see also figure \ref{fig:switchcartoon}, points A, B).

\noindent (ii). With further increase of $\mu$, the lower mean-field level 
$E_2$ crosses the Fermi level at $\mu=0.0052t$, resulting in the usual feature
(shoulder) in $\Theta_{tr}$. While $n_2(\mu)$ grows apace with $\Theta_{tr}$,
the value of $n_1(\mu)$ remains low, passing through a maximum of $n_1 \approx
0.017$
at $\mu \approx -0.01 t $. The transmission zero $Z$ remains well above 
the Fermi level, 
$Z>E_1>\mu$ (figure \ref{fig:switchcartoon}, points B, C). 
At $\mu > 0.0075 t$, two additional solutions to the mean field 
equations appear [dotted lines in figures \ref{fig:switching} {\it (a)} 
and {\it (b)}]. At the beginning, these new solutions are characterised by  
higher values of the total energy, $\Omega(\mu)$, which can be evaluated 
via equation (\ref{eq:omegatot}). 

\noindent (iii). As can be anticipated by considering the filled dot case at 
$\mu\gg E_{1,2} \approx E^{(0)}_{1,2}+U$, with increasing $\mu$ the energy 
levels $E_{1,2}$ must
eventually cross again, reverting to their original order (cf. figure 
\ref{fig:switchcartoon}, point E). In the present case,
this second crossing occurs in a discontinuous manner, {i.e.,} at some point
it becomes energetically favourable to abruptly depopulate level 2 (which then
shifts upwards) while putting most of the carrier population of the dot into 
level 1 (which is lowered). This is the {\it ``population switching'' in its 
discontinuous form} as proposed in reference \cite{SI}. Mathematically, with  
increasing  $\mu$ within the 
multiple-solution region we eventually reach the point 
$\mu_c \approx 0.0275 t$, 
where the total energy values of the two lower-energy solutions cross. Hence
a switch of the solution branch occurs, accompanied by a discontinuous change
in all the QD properties (vertical solid lines in figure \ref{fig:switching}).
While on approach to this critical value of $\mu$ from below the Fermi level
was located below the transmission zero, the situation immediately
following the transition
is that of $Z<E_1<\mu<E_2$, {\it i.e.}, the Fermi level is above the point
of transmission phase lapse. The actual point $Z=\mu$ (and the associated
phase lapse of $-\pi$), however, is never crossed because it lies
in the thermodynamically unstable region. This is illustrated by the dotted
line showing transmission phases for unstable solutions in figure 
\ref{fig:switching} {\it (b)}, which includes a phase lapse of $-\pi$ at
$\mu \approx 0.038 t$. The actual change of transmission phase at $\mu=\mu_c$,
\begin{equation}
\Delta \Theta_{tr} = -\pi + \pi \Delta N_{dot}
\label{eq:lapserenorm}
\end{equation}
includes both the contribution of this phase lapse and another term, related
by the Friedel sum rule \cite{prb06} to the jump of the ``dot-related'' 
particle number, $N_{dot} = N(\mu)-N_{w}(\mu)$. Here, $N(\mu)$ is the
total number of particles in the system, whereas $N_w$ is the number of 
particles in an unperturbed (connected) wire, equation (\ref{eq:Hwire}).
Thus, $N_{dot}$ includes both the dot occupancy, $n_1+n_2$, and the 
``dot-induced'' change of population within the leads. $N_{dot}$ must always
increase with decreasing gate voltage, or equivalently with increasing $\mu$;
in particular, it has a positive jump at $\mu=\mu_c$, renormalising
the value of phase jump $\Delta \Theta_{tr}$, equation (\ref{eq:lapserenorm}).
In our case, $\Delta \Theta_{tr} \approx  -2.514$, 
hence $\Delta N_{dot} \approx 0.20$
[note the difference of the latter from the jump in
the dot level occupancy, $\Delta (n_1+n_2)\approx 0.28$].
In the schematic representation on figure \ref{fig:switchcartoon}, the 
discontinuity (bold vertical line) occurs between the points C and D.

\noindent (iv) With further increasing $\mu$, the (narrower) dot level 1 
remains nearly filled, although its occupancy $n_1(\mu)$ does not vary
monotonously, passing through a local minimum of $n_1\approx 0.987 $ at 
$\mu \approx 0.12 t$; the corresponding energy level $E_1$ lies below the
Fermi level. The occupancy of the other level, $n_2$, increases with $\mu$,
with the level $E_2$ crossing the chemical potential at $\mu \approx 0.103 t$,
resulting in another smooth increase of $\Theta_{tr}$. We note that the same
(broad) level $E_2$ crossing the Fermi level {\it twice} 
(above and below the jump) is a known feature (``hovering level'') of the
population switching scenario \cite{Baltin,SI}. The unstable
solutions (dotted lines) provide a continuous connexion between the states 
of the system above and below the jump, with the level $E_1$ crossing the 
Fermi level at $\mu \approx 0.089 t$ (in the unstable region). With increasing
$\mu$, the pair of unstable solutions finally disappears at 
$\mu \approx 0.092 t$.  In figure \ref{fig:switchcartoon}, the second level 
crossing and the disappearance of the unstable solution occur between points 
D and E.
 
The origins of discontinuity as found at $\mu=\mu_c$ become clear if one 
considers the case when one of the QD levels is fully decoupled from
the rest of the system, {\it e. g.}, $a=0$ (corresponding to $\alpha=-1$) and
$w_0=0$. With increasing $\mu$,
the occupancy of the QD level 1 then changes abruptly form $n_1=0$ to
$n_1=1$, at which point all other QD properties 
($n_2$, $E_{1,2}$, $\Theta_{tr}$) must suffer a jump as well. In the language
of mean field (which is exact in this case as there are no quantum fluctuations
of $n_1$), this means the presence of  multiple solutions in a finite
region of values of $\mu$ near the
jump point \cite{SI}.  Indeed, assuming that  the QD energy scales
$E_2^{(0)}-E_1^{(0)}$, $b^2/t$, and $U$ are all much smaller than the 
bandwidth, $2t$, we find that the value of $E_1(\mu)$ is determined by a single
mean field equation,
\begin{equation}
E_1=E_1^{(0)}+U \left\{\frac{1}{2}+ \frac{1}{\pi} {\rm arc tan} \left[
\frac{\pi \nu_0}{b^2} \left(E_2^{(0)}+U \theta (\mu-E_1) -\mu \right)
\right] \right\}\,,
\end{equation}
where $\nu_0$ is the per-site density of states in the leads. It is easy
to verify that there is at least one multiple solution region,
where the system switches from the $E_1>\mu$ to $E_1<\mu$ branch.
These two branches of $E_1(\mu)$, connected by a segment of the singular line 
$E_1(\mu)=\mu$
(where in the $a=0$ case the value of $n_1$ is ill-defined) together form a 
z-shaped structure
similar to that shown in figure \ref{fig:switching} (or, in a cartoon
form, in figures \ref{fig:switchcartoon}  and \ref{fig:plot} below). 
When the value of 
$\alpha$ is increased
from $\alpha=-1$ (corresponding to $a \neq 0$), this picture changes 
in a continuous fashion, so that in order to eliminate the multiple
solution region (and hence the jump) altogether, $a$ must exceed a certain 
finite value, $a>a_0(U)$. In the opposite case of $a<a_0$, the discontinuity
persists, as exemplified by the $\alpha=-0.6$, $U=0.1t$ case shown in figure
\ref{fig:switching} (black lines) and discussed above.

If in the latter case of $\alpha=-0.6$, $U=0.1t$ the value of $\alpha$ is 
increased further, one finds that the while the absolute value of the jump
decreases (and the phase lapse value approaches $-\pi$), the location of 
the discontinuity shifts further to the left. By the time the discontinuity
disappears at $\alpha_c (U) \approx .012$, the phase lapse is outside the 
interval
of the values of $\mu$ between the two transmission peaks. This situation
changes when the value of $U$ is smaller; we will now turn to the $\alpha =
-0.25$, $U=0.03t$ case, shown in figure \ref{fig:switching} by the green
lines. 

In this case, the mean field energy levels $E_1$ and $E_2$ [green dashed and 
dashed-dotted lines in figure \ref{fig:switching} {\it (a)}] do not cross, 
$E_1<E_2$ for all values of $\mu$. Nevertheless, they do come close to each 
other at $\mu \approx -0.0025 t$, where $E_2-E_1 \approx 0.0014t$; a slight
increase in $|\alpha|$ would give rise to a pair of crossings, $E_1=E_2>\mu$,
in this region, without changing the overall picture. Throughout this
low-$\mu$ region (where the electron population on the dot, 
$n_1+n_2$, is below $0.5$), the occupancy $n_2$ of the broader level 2 [green 
dashed-dotted line in figure \ref{fig:switching} {\it (b)}] is larger than
that of site 1 (green dashed line). The value of $E_1$ crosses $\mu$ at
$\mu \approx  0.007t$; this is accompanied by a rapid increase in the
value of $n_1$, which exceeds $n_2$ for $\mu > 0.05t$, giving rise to a
sharp upturn in $E_2(\mu)$ [cf. equation (\ref{eq:mfee2})]. Around this
point, it becomes favourable to  depopulate level 2 (hence a downturn
in $n_2$) while increasing $n_1$. This process however happens continuously,
exemplifying the scenario of {\it continuous population switching}, as
encountered earlier in references \cite{YG04,Sindel}. Comparing this scenario
with that of $U=0.1t$, $\alpha=-0.6$ discussed above, one may say that
a slight non-monotonicity of $n_{1,2}(\mu)$ noted in that case developed 
presently into
the sharp maximum of $n_2$ at $\mu\approx 0.04 t$ and absorbed the jump in
the level occupancies.
 
With a further increase of $\mu$, level 1 remains nearly filled, whereas
$n_2$ eventually starts to grow again, with $E_2$ crossing the Fermi level
around $\mu \approx 0.03t$. This is accompanied by an increase in $E_1$,
with the difference $E_2-E_1$ approaching its bare value,  
$E_2^{(0)}-E^{(0)}_1=0.004 t$, as the chemical potential increases beyond
the dot energy range.

Throughout the entire range of values of $\mu$, the transmission zero is
located below the lower dot level, $Z<E_1$. We note, however, that
the points $\mu=\mu_{1,2}$ as defined by 
equation (\ref{eq:intdotpeaks}) (which we use as transmission peak locations,
see below), are
shifted with respect to those of the dot levels. This shift becomes
relatively more pronounced when the dot levels lie close to each other
(on the scale of the level widths), and if the value of $\alpha$ is not 
too large, the phase lapse may fall in between the two transmission ``peaks''
while remaining outside the interval between the two dot levels. This 
situation is realised
in the present case, where the transmission phase [solid green line in
figure \ref{fig:switching} {\it (b)}] suffers a lapse of $-\pi$ at
$\mu \approx 0.005t$ (the point where $n_1=n_2$), shortly above the
point of $\mu=\mu_1 \approx 0.002t$. We note a strong
asymmetry of the ``peak'', reflected in a rather irregular profile of 
$\Theta_{tr} (\mu)$ near the phase lapse. The rapid non-monotonous variation
of the dot parameters, combined with a rather small inter-level distance
$E_2-E_1$, results also in the absence of a well-defined ``shoulder''
of $\Theta_{tr} (\mu)$ associated with the second transmission peak; the value
of $\mu_2$ is about $0.034 t$. 

\begin{figure}
\includegraphics{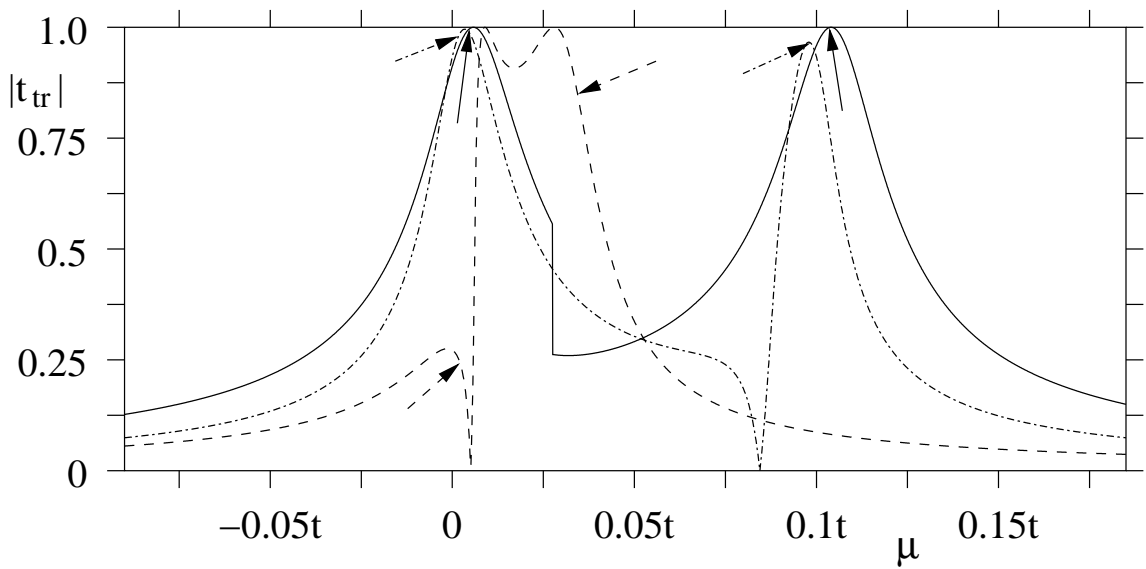}
\caption{\label{fig:trans} Transmission amplitude $|t_{tr}$ for an interacting
two-level QD as a function of chemical potential: solid line, $\alpha=-0.6$,
$U=0.1 t$, and $w_0=0$; dashed line, $\alpha = -0.25$, $U=0.03 t$, and $w_0=0$;
dashed-dotted line, $\alpha=0.3$, $U=0.1t$, and $\kappa=0.35$ (corresponding
to $w_0=0.0015t$). The rest of parameters for all three cases are $\tilde{E}_1^{(0)}=0$, $\tilde{E}_2^{(0)}=0.004t$, and $\sqrt{a^2+b^2}=0.125t$. The arrows 
show the locations of the corresponding transmission peaks as given 
by equation 
(\ref{eq:intdotpeaks}).}
\end{figure}

These features are further illustrated by a plot of the transmission
amplitude $|t_{tr}(\mu)|$,
\begin{eqnarray}
\fl
|t_{tr}|^2=&&{(t^2-\mu^2) 
\left[\frac{b^2}{t^2}(\tilde{E}_1-\mu)-\frac{a^2}{t^2} (\tilde{E}_2-\mu) 
\right]^2}
\left\{\left[\frac{a^2}{t^2}(\tilde{E}_2-\mu)
+\frac{b^2}{t^2} (\tilde{E}_1-\mu) +2\mu \frac{a^2b^2}{t^4} \right]^2 \times 
\right. \nonumber \\
\fl
&&\left.
\times (t^2-\mu^2)
+\left[(\tilde{E}_1-\mu+\frac{a^2}{t^2}\mu)(\tilde{E}_2-\mu+\frac{b^2}{t^2}\mu)
-(t^2-\mu^2)\frac{a^2b^2}{t^4} -\frac{1}{4}w^2\right]^2\right\}^{-1}
\label{eq:intdottrans}
\end{eqnarray}
as a function of $\mu$ (figure \ref{fig:trans}), where the case of $U=0.03t$,
$\alpha=-0.25$ is represented by the dashed line. We see that in this case,
$|t_{tr}|$ does not show the two well-separated peaks as anticipated in the
Coulomb blockade regime. This is due  to a relatively small value of $U$
(and hence the small mean-field level separation). We also note that
the profile of transmission is rather irregular.  The points marked by the
dashed arrows [locations of $\mu_{1,2}$ as given by equation 
(\ref{eq:intdotpeaks})] do not correspond to any particular features of the
plot. This is not surprising, since these values of $\mu=\mu_{1,2}$ correspond
to half-integer values of $N_{dot}$ and should approach the transmission
peak locations  in the Coulomb blockade regime only.
  
We note the presence of a transmission zero, corresponding to a phase
lapse; this is surrounded by the two peaks reminiscent of ``correlation
induced resonances'' which were first reported in reference \cite{Marquardt}
and subsequently explained by the Kondo-type physics \cite{Lee06,Avi}. 

The solid line in figure \ref{fig:trans} corresponds to the discontinuous 
population switching case of $U=0.1t$ and $\alpha=-0.6$. We see the two 
well-separated Coulomb blockade peaks of roughly the same width, whose maxima
lie very close to the values of $\mu_{1,2}$, equation (\ref{eq:intdotzero}).
As expected \cite{prb06,Goldstein}, there is no transmission zero, 
since the actual $t_{tr}(\mu)=0$
point belongs to the unstable solution. The discontinuity is clearly seen on 
the plot. An interesting feature of both dashed
and solid curves in figure \ref{fig:trans} is the presence of two maxima where
the value of $|t_{tr}(\mu)|$ exactly equals one. An investigation of
equation (\ref{eq:intdottrans}) suggests that this is always the case for the 
symmetric situation ($w_0=0$), provided that the mean field value of
$|\tilde{E}_2-\tilde{E}_1|$ (which in the partially filled regime is of
the order of $U$) is larger than $|2ab/t|$ (here, we assume that both 
$|E_{1}-E_2|$ and $(a^2+b^2)/t$ are much smaller than the bandwidth of the 
leads, $2t$).

This situation changes in the asymmetric, $w_0\neq 0$ case, as exemplified
by the dashed dotted line in figure \ref{fig:trans}. This line, which 
illustrates the effect of interaction-induced ``excitonic correlations'' (as 
discussed in section
\ref{sec:exciton} above) shows two clear Coulomb blockade peaks of unequal
height, with a transmission zero in between. Again, equation 
(\ref{eq:intdotpeaks}) gives accurate values for the peak locations.

~\\
\centerline{* * *}

We are now in a position to summarise the behaviour of an interacting two-level
dot in the three regimes considered thus far:

\noindent (a). Right-left asymmetry ($w_0\neq 0$), excitonic correlations 
(section \ref{sec:exciton}). Within ``phase'' 1 of figure \ref{fig:phadiag},
this yields a phase lapse of $-\pi$ between the two transmission peaks.
The underlying mechanism is the interaction-induced change in the sign
of the dot-lead coupling, the
evolution of dot parameters with varying $\mu$ is continuous throughout,
the transmission peaks are of unequal height and are separated by a
transmission zero.    

\noindent (b). Right-left symmetric ($w=0$) case showing a discontinuous
QD level population switching as a function of $\mu$. The locus of the 
corresponding points
in figure \ref{fig:phadiag} is at the lower edge of the green area marked
as ``phase'' 3. While the interaction-induced sign change is impossible,
the phase lapse is located between the two transmission peaks of height
$|t_{tr}|=1$, due to
the re-ordering of respective locations of the peaks and the transmission
zero at the point of discontinuity. The latter coincides with the phase
lapse (whose magnitude is however reduced to a quantity between $-\pi$ and 0)
and with the transmission minimum (where the transmission amplitude
retains a finite value).

\noindent (c) Continuous population switching. This case would appear
at a ``phase diagram'' for a lower value of $U$ as the interval of
the $\kappa=0$ axis where the blue region of ``phase 1'' extends all
the way down to this axis. While the example we considered lies outside
the Coulomb blockade region, one observes a phase lapse of $\pi$, coinciding
with a transmission zero and surrounded by the ``peaks'' of transmission.
This mechanism is driven by a strong non-monotonicity of the QD level
occupancies in this regime.

Scenarios (a) and (c) are both continuous and can be expected to be robust
with respect to quantum fluctuations (not included in the present treatment).
This may not be the case for the discontinuous scenario (b), which is therefore
likely to be partly replaced with scenario (c). 

We now turn to the intermediate values of $\kappa$ in order to understand the
structure of the entire ``phase diagram'', figure \ref{fig:phadiag}, in terms
of interpolation between these three regimes. 

\section{Interplay between the 
Two Mechanisms: The Mean Field ``Phase Diagram''}
\label{sec:phases}  

{\it Minimal value of 1-2 asymmetry required for the correlation-induced
switching in the phase lapse location. Mean field ``phase diagram'' as a
result of superposition of the two correlated mechanisms: descriptions
of different ``phases''. ``Phase diagram'' in the case of a smaller $U$ and
continuous population switching}.
\\

In this section, we turn to the intermediate values of $\kappa$ [equation
(\ref{eq:phadiagparameters})] in order 
to understand the
structure of the entire ``phase diagram'', figure \ref{fig:phadiag}, in terms
of interpolation between the different manifestations of interaction-induced
correlations considered above -- effective sign change
due to excitonic correlations at larger $\kappa$ (section \ref{sec:exciton}) 
vs. population switching at 
$\kappa=0$ (section \ref{sec:switching}). 

Both of these mechanisms result in the occurrence of a phase
lapse between the two transmission peaks, and the underlying physics is
partially similar in that once the chemical potential lies within a (broadened)
dot level, the energetically preferable situation corresponds to the broader
of the two QD levels being partially filled. This in turn implies that
on approach of the value of $\mu$ to the dot level energies from the side of
the weaker-coupled (bare) level ({\it i.e.,} $\mu<E_{1}^{(0)}<E_2^{(0)}$ 
for $\alpha<0$), 
a level inversion must take place, with the broader level approaching the
chemical potential first. At $\kappa=0$ (no hybridisation) this inversion
takes the form of an actual level crossing, whereas at $\kappa>0$ only the
``site energies'' can cross,
\begin{equation}
\tilde{E}_1(\mu_c)=\tilde{E}_2 (\mu_c)
\label{eq:sitescross}
\end{equation}
(figure \ref{fig:switchcartoon}, points A and B).
Generally,
the occurrence of the crossing, equation (\ref{eq:sitescross}), indicates
that the correlation effects are sufficiently strong to activate at least one
of the mechanisms responsible for the occurrence of the phase lapse between
the transmission peaks.
      In the case of chemical potential lying well below the dot energy
levels, the two site occupancies are given by
\begin{equation}
\tilde{n}_1 \approx \frac{a^2\sqrt{t^2-\mu^2}}{\pi t^2(\tilde{E}_1-\mu)}\,,\,\,\,\,\,\tilde{n}_2\approx \frac{b^2\sqrt{t^2-\mu^2}}{\pi t^2(\tilde{E}_2-\mu)}
\label{eq:lowocc} 
\end{equation}
[see reference \cite{prb06}, equations (16-17)]. In writing  equation 
(\ref{eq:lowocc}) we assumed that $t \gg \tilde{E}_{1,2} -\mu$ [which allows
to use the constant value $\nu_0=1/(\pi \sqrt{t^2-\mu^2})$ for the density
of states in the leads\footnote{The dependence of $\nu_0$ on energy is indeed
a weak effect, accounting for a slight asymmetry between $\alpha>0$ and $\alpha
<0$ cases in figure \ref{fig:phadiag}.}] 
and $\tilde{E}_{1,2} -\mu\gg a^2/t, b^2/t$. In
addition, the value of $w$ [equation (\ref{eq:mfew})] should not be too large, 
$w^2 \ll a^2(t^2-\mu^2)/t^2$ and $w^2 \ll b^2(t^2-\mu^2)/t^2$.  Since in this
range of values of chemical potential $w$ stays close to its bare value,
$w \approx w_0$,  [cf. figure \ref{fig:exciton}] 
the latter is not a restrictive condition. The site energies 
crossing may occur below
the transmission peaks only in the $a<b$ ($\alpha < 0$) case, which we will
consider here. Then, equations (\ref{eq:mfee2}) yield the following
condition for the crossing point $\mu_c$:
\begin{equation}
\tilde{E}_{1,2}(\mu_c)- \mu_c =\frac{U \sqrt{t^2-\mu_c^2}}{\pi t^2} 
\frac{b^2-a^2}
{\tilde{E}_2^{(0)}-\tilde{E}_1^{(0)}}\,.
\end{equation}
The two transmission peaks are well defined only if the dot levels are well
separated at each peak. Hence the crossing point $\mu_c$ must lie well below
the lowest peak, in the region where the net occupancy of the dot, 
$\tilde{n}_1+\tilde{n}_2$, is small:
\begin{equation}
\tilde{E}_{1,2}(\mu_c)- \mu_c \gg 
\Gamma_1+\Gamma_2=\frac{a^2+b^2}{t^2}\sqrt{t^2-\mu_c^2}\,,
\label{eq:strongerU}
\end{equation}
where the quantity on the r. h. s. is the combined 
broadening of the two levels, cf. equation
(\ref{eq:broadening}). We note that the condition (\ref{eq:strongerU}) is also
required for equations (\ref{eq:lowocc}) to hold. In terms of our 1-2 coupling
asymmetry parameter $\alpha$ [see equation (\ref{eq:phadiagparameters})], 
equation (\ref{eq:strongerU}) takes form 
\begin{equation}
U \gg U_c \approx \frac{\pi \left( \tilde{E}_2^{(0)}-\tilde{E}_1^{(0)}\right)}
{|\alpha| \sqrt{2-\alpha^2}}\,,
\label{eq:Uc}
\end{equation}
where we wrote $|\alpha|$ instead of $\alpha$ in order to include also the 
$a>b$ case, when the point $\mu_c$ lies above the transmission peaks. The
value of $U_c$ sets the scale for the interaction strength required to
produce strong correlation effects. At a fixed $U$, these effects can be
amplified by increasing the level coupling asymmetry so that
\begin{equation}
1-\alpha^2 \ll 1- \alpha_c^2 \approx \sqrt{1-\frac{\pi^2 \left( 
\tilde{E}_2^{(0)}-\tilde{E}_1^{(0)}\right)^2}{U^2}}\,.
\label{eq:alphac}
\end{equation} 
For the values of parameters used in figure \ref{fig:phadiag} this yields
$\alpha_c \approx 0.09$, which is a fairly accurate estimate for the width
of the red area, occupied by ``phase'' 2. We already mentioned that in
this ``phase'' the phase lapse occurs outside the energy interval between
the two transmission peaks, due to insufficiently strong correlation
effects. 

\begin{figure}
\includegraphics{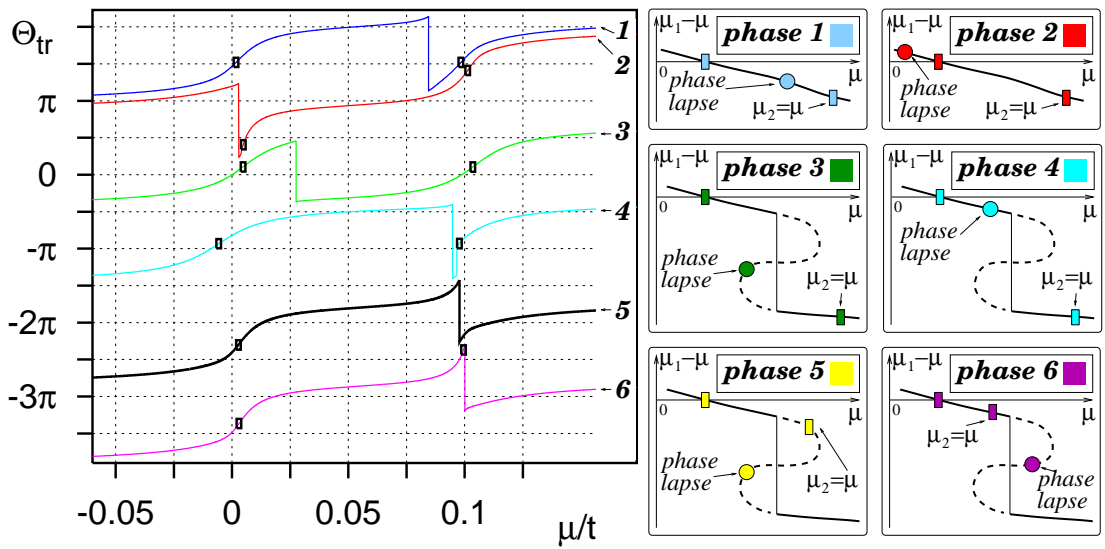}
\caption{\label{fig:plot} (colour) Typical behaviour of
$\Theta_{tr}(\mu)$ in different ``phases'' ( left; plots shifted vertically
for convenience). Relative positions of transmission peaks, $\mu_{1,2}$, 
(boxes; also in the left panel)
and the $-\pi$-phase lapses (circles) in ``phases'' 1-6 are clarified by the 
schematic plots of $\mu_1-\mu$ around the multiple-solution
region (absent for ``phases'' 1-2). Solid (dashed) lines correspond
to stable (unstable) solutions. The abrupt ``switching'' of
solutions (vertical solid line) may either renormalise the phase lapse
 (when the $- \pi$-lapse lies in the unstable region) or result in a
positive jump of $\Theta_{tr}$.}
\end{figure}

We will now turn to figure \ref{fig:phadiag} and summarise the mean-field
properties of QD within the parameter region corresponding to each ``phase''.
Typical $\Theta_{tr}(\mu)$ profiles for each of the ``phases'' are shown in the
main panel of figure \ref{fig:plot}.

\noindent {\bf ``Phase'' 1} (blue). In this ``phase'', a phase lapse of $-\pi$
occurs between the two transmission peaks [marked by the two boxes on the 
$\Theta_{tr}(\mu)$ plot]. There is no discontinuity in the evolution of
other QD parameters with varying $\mu$, as shown in figure \ref{fig:plot}
by a schematic $\mu_1(\mu)-\mu$ vs. $\mu$ plot which does not have a multiple-
solution region. As explained in section \ref{sec:exciton}, the underlying
mechanism is that of an effective coupling sign change due to excitonic
correlations, and the corresponding region is in the upper part of the
``phase diagram'' (larger $\kappa$), away from the $\alpha=0$ axis.


\noindent {\bf ``Phase'' 2} (red). A phase lapse of $-\pi$ occurs outside
the interval between the two level crossings, $\mu_{1,2}(\mu)=\mu$. The
corresponding area is a narrow stripe around the $\alpha=0$ axis, and at
small or moderate values of $\kappa$ its width can be estimated with
the help of Eq. (\ref{eq:alphac}). Owing to an insufficient 1-2 level
coupling asymmetry, correlation effects are not strong enough to cause
either a sign change or a discontinuity (note the absence of a multiple
solution region in a schematic plot of $\mu_1(\mu)-\mu$). Away from
the $\alpha=0$ axis the ``site occupancies'' $\tilde{n}_{1,2}(\mu)$ may
still show a non-monotonous behaviour and continuous population switching
(with the phase lapse located away from the transmission peaks). 

\noindent {\bf ``Phase'' 3} (green). In this case, a ``renormalised'' phase
lapse, $\Delta \Theta_{tr}>-\pi$, occurs between the two transmission peaks.
This ``phase'' occupies the area adjacent to the $\kappa=0$ axis, excluding
the vicinity of the fully symmetric case, $\kappa=\alpha=0$. 
The underlying mechanism (discontinuous population switching) was discussed
in section \ref{sec:switching}. Note the presence of a ``fold'' on the
schematic plot of $\mu_1(\mu)-\mu$, indicating the presence of a multiple
solution region. The unstable part of the solution is shown by the dashed 
line and includes the transmission zero (circle), whereas the vertical
solid line corresponds to a discontinuous jump between the two stable branches.
Since at $\kappa=0$ the multiple solution region has a finite width, the
effects of right-left asymmetry (favouring the change of the coupling sign)
cannot eliminate the discontinuity also at small but nonzero values of 
$\kappa$, corresponding to this ``phase''.

The case of a fully decoupled level $\alpha=\pm1$, $\kappa=0$ is essentially 
the one considered in the reference \cite{SI} for a multi-level QD. We note 
that the value of
phase lapse at these two points also differs from $-\pi$, contrary to
earlier expectations \cite{SI}. For the values of the QD parameters used in
figure \ref{fig:phadiag}, we find $\Delta \Theta_{tr} \approx - 2.4$ at 
$\kappa=0$, $\alpha=-1$, and $\Delta \Theta_{tr} \approx - 2.5$ at 
$\kappa=0$, $\alpha=1$.

\noindent {\bf ``Phase'' 4} (cyan). This ``phase'' lies below the bold line 
which in figure \ref{fig:phadiag} forms the lower boundary of the ``phase'' 
1 area and  corresponds to the
onset of discontinuity in the evolution of the dot parameters with $\mu$.
With decreasing $\kappa$, this onset occurs in a continuous fashion, via the
point where at some value of $\mu_0$ the derivative of the QD parameters
[such as $d \mu_1(\mu)/d\mu$] becomes infinite. With a further decrease
of $\kappa$ this derivative changes sign [in the case of $d \mu_1(\mu)/d\mu$,
becomes positive], and a multiple solutions area forms around $\mu_0$.
Initially, this area is narrow and the unstable solution includes neither 
transmission zero nor transmission peaks (since in general $\mu_0$ dose not 
correspond
to either of those). This is represented by the corresponding schematic
plot of $\mu_1(\mu)-\mu$ in figure \ref{fig:plot}; the jump between the
two branches yields a discontinuous increase of the transmission phase, 
$\Delta \Theta_{tr}^{(2)}>0$, which is clearly visible in the 
$\Theta_{tr}(\mu)$ plot\footnote{When plotting $\Theta_{tr}(\mu)$ in this
case, we used different values of the QD parameters in order to make
this behaviour more pronounced.} to the right of the transmission zero, 
$\Delta \Theta_{tr}^{(1)}=-\pi$. When the value of $\kappa$ is decreased 
further,
the unstable part of solution spreads to include the transmission zero,
and we find ourselves within the ``phase'' 3. Since ``phase'' 4 interpolates 
between ``phases'' 1 and 3, it is clear that both phase jumps lie in 
the region between
the two transmission peaks.

\noindent {\bf ``Phase'' 5} (yellow). This ``phase'' occupies the areas 
adjacent to those of ``phase'' 3 from the side of smaller values of the
1-2 level coupling asymmetry, $|\alpha|$. It lies in the discontinuous
region of the ``phase diagram'' below the bold solid line.
  The decrease of $|\alpha|$ within ``phase'' 3 causes the discontinuity (which
in this case includes a jump ``over'' the transmission zero) to shift towards
one of the transmission peaks. Eventually the unstable area moves to include
this peak as well, as can be seen from the schematic plot of $\mu_1(\mu)-\mu$.
Thus the transmission peak is circumvented by a jump, as reflected by the
absence of the corresponding box (and the corresponding inflexion point)
on the $\Theta_{tr}(\mu)$ plot, which also includes a single renormalised
phase lapse,  $\Delta \Theta_{tr}>-\pi$.

\noindent {\bf ``Phase'' 6} (magenta). When the value of $|\alpha|$ in
``phase'' 5 is decreased further, the multiple solution area moves further
away from the centre of the energy interval between the two transmission peaks.
Eventually the unstable part (which still includes the transmission zero)
clears this interval altogether, restoring the transmission peak to the 
stable branch. This corresponds to crossing the boundary from ``phase'' 5 into
``phase'' 6. The renormalised phase lapse, $\Delta \Theta_{tr}>-\pi$,
is then located outside the area between the two peaks.

\noindent {\bf ``Phase'' 7} (not shown due to the small area it occupies).
With a further decrease of $|\alpha|$, the multiple solution area shrinks
(while shifting away from the transmission peaks), and the transmission
zero returns to a stable branch. The result is similar to ``phase'' 4 
above, with the only difference that the two phase jumps now lie outside
the region between the two peaks.

A further decrease in $|\alpha|$ results in the disappearance of the multiple
solution region and crossing the bold line into ``phase'' 2 so that the
sequence of ``phases'' 5, 6, and 7 interpolates between ``phases'' 3 and 2.
The seven ``phases'' discussed above do not
exhaust all the possibilities of mutual overlaps between the multiple solution
region of values of $\mu$ (and associated discontinuity), transmission zeroes,
and transmission peaks. Some of the other ``phases'' occupy minute areas 
(not shown) near the crossing of the boundary between ``phases'' 1 and 2
and the bold line;  still others do not arise 
for the values of parameters used in figure \ref{fig:phadiag}.


We see that the whole of ``phase diagram'', figure \ref{fig:phadiag} can be 
understood in terms of interpolation between the three cases: that 
of ``phases'' 2 (where the interaction does not affect the location
of the phase lapse between the transmission peaks), 3 (discontinuous population
switching), and 1 (effective coupling sign change). Of these, the latter
occupies the largest area. We will discuss the implications of this
findings in the following section.  

It should be emphasised  that in figure 
\ref{fig:phadiag} the boundaries between ``phases'' 1 and 2,
and between ``phases'' 3, 5, and 6  are a matter of convention and do not 
correspond to sharp transitions of any kind. Rather, they merely mark the 
changes in mutual locations of transmission peaks {\it as defined by equation
(\ref{eq:intdotpeaks}) } on one hand, and phase lapse/discontinuity on the
other. 

When the value of $U$ is reduced to $U<0.04t$ (while keeping the other QD
parameter values in figure \ref{fig:phadiag} constant), the bold line 
marking the onset of the discontinuity no longer intersects the 1-2 ``phase''
boundary. In this case, ``phases'' 5, 6, and 7 diasppear and the area of
``phase'' 1 formally extends down to the $\kappa=0$ axis at a certain
range of values of $\alpha$. In the $U=0.03t$ case [see the green lines
in figure \ref{fig:switching}], this range is given by $\alpha^{(-)}_1 \approx
-0.31 < \alpha < \alpha^{(-)}_2 \approx -0.2$ and $ \alpha^{(+)}_2 \approx
0.18 < \alpha <  \alpha^{(+)}_1 \approx 0.3$. At this values of $\alpha$ and
for small right-left asymmetry $\kappa \ll 1$,  the ``continuous population 
switching'' scenario 
as discussed in section \ref{sec:switching}  
{\it formally} results in an occurrence of the phase lapse between the two
transmission peaks [as defined by equation (\ref{eq:intdotpeaks})]. 
We stress however that at least for the moderate values of $\kappa$, 
the entire region, $\alpha^{(-)}_1 < \alpha < \alpha_1^{(+)}$ the QD is outside
the Coulomb blockade regime (transmission peaks are not well separated, as
exemplified by the dashed line in figure \ref{fig:trans}). 
As a consequence, equation (\ref{eq:intdotpeaks}) 
(used by us to define ``phase''1) looses accuracy, making our convention for
distinguishing between ``phases'' 1 and 2 problematic.

We note that the values of $\alpha^{(\pm)}_1$, marking the onset of the
discontinuous behaviour, are in a good agreement with equation 
(\ref{eq:alphac}),
which at $U=0.3t$ yields $|\alpha_c| \approx 0.3$. With a further increase of
$|\alpha|$ towards $|\alpha|=1$ the Coulomb blockade effects set in even
at $U=0.3t$.

\section{Discussion}
\label{sec:conclu}

{\it Conclusions of the present mean-field study. Anticipated results
for a mean-field treatment of multi-level dots. Role of quantum 
fluctuations: available results and outstanding questions.}
\\

Broadly speaking, our mean-field treatment of the interacting two-level dot
yields the following result, which holds provided that the interaction is
sufficiently strong, see equation (\ref{eq:Uc}): 

\noindent {\it Irrespective of the original sign of the dot-lead coupling,
the transmission phase lapse in an interacting QD generally occurs between
the two transmission peaks}.

There are {\it two distinct correlation-induced mechanisms} which bring about
this uniform situation. Of these, one is related to the off-diagonal 
(``excitonic'') correlations on the dot \cite{prb06} and requires the presence 
of a finite left-right asymmetry in the original dot-lead coupling (as can be 
expected generally in the experimental realisations). The other, which in its
pure form is operational in the left-right symmetric case, has to do with 
the correlation-induced ``population switching'' \cite{Baltin,SI}, which
can occur either discontinuously (in the Coulomb blockade regime) or
continuously (this leads to the Coulomb blockade being lifted). 

At the most
basic level, the two mechanisms share the same origin, familiar from
the standard solid state physics: namely, when a band
or an impurity level resides at the chemical potential and is therefore
partially filled, energy may be gained by increasing the width of the band
or by broadening  the impurity level. In case of the ``excitonic'' 
mechanism, this ``broadening'' of the narrower QD level is achieved via a 
strong increase of hybrydisation with the broader level. This in turn is 
associated with the change of the coupling sign, causing the $-\pi$ phase 
lapse to occur between the two transmission peaks. In case of the 
discontinuous population switching
\cite{SI}, the levels of a partially filled dot are abruptly rearranged in
such a way that the narrower level never actually crosses the chemical 
potential. This ``jump'' of the narrow level (say, $E_1$) from $E_1>\mu$ to
$E_1<\mu$ with increasing $\mu$ is accompanied by a similar jump of the
transmission zero $Z$ and therefore leads to an abrupt change of a transmission
phase, $\Delta \Theta_{tr}>-\pi$ (renormalised phase lapse \cite{prb06}).
Continuous population switching in the relevant regime (which arises only 
for smaller values of $U$, see sections \ref{sec:switching} and 
\ref{sec:phases}) involves the two levels approaching the
chemical potential at the same time (along with the transmission zero) and
is accompanied by a phase lapse of $-\pi$ in the absence of the two 
well-defined transmission peaks. 
  
The behaviour of a QD for general values of the left-right and 1-2 coupling
asymmetries can be understood in terms of {\it superposition of these two 
mechanisms}. The continuous population switching scenario can evolve
into the ``excitonic'' one via a smooth crossover with increasing left-right
asymmetry. The interplay of these two with the 
discontinuous population switching mechanism, on the other hand, gives rise to
a number of different intermediate ``phases''.  These are characterised by
different numbers (1 or 2) and mutual locations of phase jump(s) and 
transmission peak(s). Our results suggest that on the whole, 
the behaviour found in most cases is the one dominated by the ``excitonic''
mechanism, with a phase lapse of $-\pi$ located between the two well
separated Coulomb blockade peaks.

From the theoretical standpoint, our results give rise to the following
two questions: (i) how are these findings generalised in the case of a
multi-level interacting QD? (ii) What is the role of fluctuations, neglected
in our mean-field treatment? We will now address these issues in some detail.

\noindent (i) {\it Multi-level dots within the mean-field approach}. 
In general, one can expect that {\it our
conclusions will hold for a multilevel dot}, with the two levels nearest the
chemical potential playing the role of an ``effective'' two level QD, which
at least at the mean field level would behave in a qualitative agreement with
our results. One expected change is that {\it the parameter area corresponding 
to 
the effective sign change (``excitonic'') mechanism will be expanded} at the
expense of the other ``phase diagram'' regions (dominated in the mean field
approach by the population switching). Indeed, for a two-level
dot the ineffectiveness of this mechanism in the absence of the right-left 
asymmetry (section \ref{sec:exciton}) can also be viewed as originating 
from the fact that each of the dot levels is coupled to a different ``subset'' 
of carriers in the lead (odd and even wavefunctions), making the inter-level
hybridisation impossible. For a multi-level dot, the number of dot levels
exceeds the number of such subsets (which of course remains equal to two, 
even and odd). Hence even if the two levels adjacent to the chemical potential
have different coupling signs and only a small right-left asymmetry, each of 
these will hybridise with other levels
further away (some hybridised combinations may even be decoupled from the
leads \cite{berkvoppen}. This will in turn give 
rise to a left-right asymmetry of the ``effective'' two-level QD  provided 
these other
levels have asymmetric couplings to the two leads. On the other case, we saw
(section \ref{sec:exciton}) that even a moderate amount of initial left-right
asymmetry is sufficient to activate the ``excitonic'' mechanism of the
effective coupling sign change. 

These conclusions are in a qualitative agreement with the numerical results
for multilevel dots \cite{Meden}. We note, however, that these calculations
\cite{Meden}
were performed using the functional renormalisation group method, and 
therefore include fluctuations at some level.

We also note that within the mean-field approach the {\it discontinuous 
evolution}
of the QD parameters (driven by the discontinuous population switching 
mechanism) {\it will persist within  the corresponding range of parameter 
values also in the case of a multi-level QD}.
The physical reason for this is the same as in the two-level case (section 
\ref{sec:switching}): suppose that one of the QD levels is fully uncoupled 
from the leads (and from the other levels). It is then clear that with 
increasing
$\mu$ it will eventually be filled in a discontinuous manner (the occupancy
jumping from 0 to 1), leading to a discontinuity in all the QD properties 
\cite{SI}. {\it Within the mean-field treatment}, such a disontinuity 
originates from a jump between different solutions to the mean field equations,
and therefore indicates
a presence of a multiple-solution area in the parameter space. This area has a
finite size, and therefore removing it altogether and thus eliminating the discontinuity requires a finite (as opposed to 
infinitesimal) dot-level coupling. Like in the two-level case, the
discontinuous evolution will be accompanied by a renormalisation \cite{prb06} 
of the 
corresponding phase lapse values, $\Delta \Theta_{tr}>-\pi$; the conductance
at the corresponding values of chemical potential or gate voltage would not 
vanish. As described in section \ref{sec:phases}, there will also arise a
number of borderline ``phases''; in some of these, the additional positive
phase jumps would appear.

\noindent (ii) {\it Role of quantum fluctuations} (validity of the 
mean field approach). The present mean field treatment does not include
the effects of fluctuations. It is therefore important to understand to
what extent do these alter the overall picture. While considerable effort
has been made recently in this direction \cite{Marquardt,Meden,vonDelft,Kim06,Lee06,
Avi}, certain questions still remain 
unanswered. Below we will attempt to summarise the available results while
pointing out the open problems.

First, we note that the available numerical and renormalisation group studies
suggest that the {\it generic effects of electron-electron interaction on the
dot include the appearance of the phase lapse between the two transmission
peaks}. This is in qualitative agreement with the mean field results.
On the other hand, the essentially many-body features like the 
``resonances'' surrounding
the transmission zero \cite{Marquardt,Meden}, which originate from the
Kondo-type physics \cite{Lee06,Avi} {\it cannot} be reproduced at 
the mean field
level.

Our mean field results suggest the {\it presence of  two distinct correlated 
mechanisms} causing the phase lapse to occur within the inter-level energy
interval. This is in line with  recent renormalisation group results 
\cite{Avi},
suggesting that the QD behaves differently in the right-left symmetric
(``parallel effective field'' \cite{Avi}) and asymmetric (``tilted effective
field'') cases. 

Of the two mechanisms identified within the present mean field approach, 
 the  {\it ``excitonic''} one (involving {\it off-diagonal
correlations on the dot}, section \ref{sec:exciton}), 
which requires the presence of some right-left asymmetry in the dot-lead
coupling ($w_0 \neq 0$), is found to be   more generic. Indeed, it gives 
rise to the
prevalent ``phase'' 1 in our ``phase diagram'', figure \ref{fig:phadiag}.
This mechanism is not related to instabilities of any kind, and can be
expected to remain robust beyond the mean field. While this is again in
line with the available results, a more qualitative comparison can and should 
be made in order to confirm that we correctly identified the underlying
physics. To this end, one should verify that the off-diagonal average, 
$\tilde{n}_{12} =\langle \tilde{d}^\dagger_1 \tilde{d}_2 \rangle$ indeed 
shows a sharp peak
in the regime of partial QD occupancy [cf. figure \ref{fig:exciton} {\it (b)}].
While the quantity $\tilde{n}_{12}$ should be readily avaliable from numerical
calculations, we are not aware of any published results for 
it\footnote{Our results are in a qualitative agreement with 
Equation (52) of reference \cite{Avi}  which suggests that 
$\tilde{n}_{12}$ reaches a maximum in the general area of the point where 
$\tilde{n}_1=\tilde{n}_2$. This result \cite{Avi} should become exact in the
large-$U$, $\tilde{n}_1+\tilde{n}_2 \rightarrow 1$ limit 
(``local moment regime'' \cite{Avi}).}.  We note that
the diagonal average values, $\tilde{n}_{1,2}$, have been
calculated recently by the numerical renormalisation group \cite{Meden} and
functional renormalisation group \cite{vonDelft} methods, as well as 
analytically \cite{Avi}.
 In the relevant range of values
of parameters, their dependence on the gate voltage shows strong 
non-monotonicity (first noticed in the mean-field studies of
the right-left symmetric case \cite{YG04,Sindel}) and looks rather similar 
to the mean field
results as seen in figure \ref{fig:exciton} {\it (b)}.
Another interesting question is related to the degree of 1-2 level coupling
asymmetry required for theis mechanism to be effective. The corresponding
mean-field result, equation (\ref{eq:alphac}), differs from the one obtained
via the renormalisation group calculation in the ``local moment'' regime 
[reference \cite{Avi}, equation (70)]
and a systematic numerical investigation is required in order to
establish whether (and when) either of these is close to the actual value. 

The other mean field mechanism which gives rise to a phase
lapse between the Coulomb blockade peaks is that of 
{\it discontinuous population
switching} (section \ref{sec:switching}). While the physical origins of the
discontinuity are quite clear (see above), the associated notion of multiple 
solution  region is restricted to the mean field approach. Thus, the
discontinuous scenario will surely be strongly affected by the fluctuations.
At present, it is not clear whether it survives in the exact treatment. 
Neither the discontinuity,
nor the associated ``renormalised'' phase lapses \cite{prb06} were  
found (except at isolated points \cite{vonDelft}) in either the functional 
renormalisation group \cite{Meden,vonDelft}, numerical renormalisation group 
\cite{Meden,vonDelft},  or analytical renormalisation
group/Bethe {\it ansatz} \cite{Lee06,Avi} treatments. 


Lastly, we wish to turn to the situation of {\it perfect right-left symmetry}
($w_0=0$) and argue that it {\it represents a singular case where the effects
of fluctuations are most pronounced}. Indeed, owing to the symmetry of this
case the hybridisation, $\tilde{n}_{12}$, and hence the effective intra-dot
hopping $w$ remain equal to zero for all values of chemical potential or
gate voltage (section \ref{sec:exciton}). Hence the transmission zero, equation
(\ref{eq:intdotzero}), remains outside the energy interval between the two
transmission peaks at all times (at least in the Coulomb blockade regime,
which is of interest to us here). We conclude that  within the {\it mean field
approach} for $w_0=0$ the only possibility to observe a phase lapse between 
the two well separated (Coulomb blockade) transmission peaks is via a 
discontinuous
restructuring of the spectrum, as explained in section \ref{sec:switching} 
(figure \ref{fig:switchcartoon}).
In this case, the phase lapse is renormalised \cite{prb06}, and is not 
accompanied by a transmission zero \cite{prb06,Goldstein}.

On the other hand, the numerical results\cite{Meden}, as well as those of
the renormalisation group approach \cite{Lee06,Avi}, reliably indicate that
at least for some values of 1-2 asymmetry in the right-left symmetric case,
the phase lapse of exactly
$-\pi$, associated with a transmission zero, does occur between the two Coulomb
blockade peaks. This cannot be reconciled with the mean field picture 
{\it at all}, via any sort of renormalisation of the mean field parameters
(which could be held responsible for the suppression of discontinuous
behaviour at $w_0 \neq 0$). Rather, fluctuations in this case must be giving 
rise to a totally new physics (albeit perhaps only in the narrow interval of 
values of chemical potential/gate voltage around the transmission zaro). 
We note that from the point of view of
renormalisation group analysis \cite{Avi} the $w_0=0$ (``parallel effective 
field'') situation does correspond to a special case, and its relationship
to behaviour at $w_0 \neq 0$ remains unclear.

\ack

The authors  thank A. Aharony, R. Berkovits, J. von Delft, 
O. Entin-Wohlman, M. Goldstein, M. Heiblum, V. Kashcheyevs, I. V. Lerner, 
V. Meden, Y.
Oreg, A. Schiller, and  H. A. Weidenm\"{u}ller for enlightening discussions. 
This work was  supported 
by the
ISF (grants No. 193/02-1, No. 63/05, and the Centers of Excellence Program), by
the BSF (grants \# 2004162 and \# 703296),
and by the Israeli Ministry of Absorption. YG was also supported by an
EPSRC fellowship.


\section*{References}

\begin{thebibliography}{99}
\bibitem{Yacoby} Yacoby A, Heiblum M, Mahalu D and  Shtrikman H 1995 
{\it Phys. Rev. Lett.} {\bf 74} 4047 
\bibitem{Schuster} Schuster R, Buks E, Heiblum M, Mahalu D, Umansky V 
and Shtrikman H 1997 {\it Nature} {\bf 385} 417 
\bibitem{Avinun} Avinun-Kalish M, Heiblum M, Zarchin O, Mahalu D
 and Umansky V 2005 {\it Nature} {\bf 436} 529
\bibitem{Ora} Entin-Wohlman O, Hartzstein C and  Imry Y 1986 {\it Phys. Rev.} B
{\bf 34} 921 
\bibitem{GIA} Gefen Y, Imry Y and  Azbel' M Ya 1984 {\it Phys. Rev. Lett.}
 {\bf 52} 129 
\bibitem{YGreview} Feldman D E and  Gefen Y 2001 unpublished 
\nonum Gefen Y 2002  {\it Quantum Interferometry with Electrons:
Outstanding Challenges} ed Lerner I V {\it et al} 
(Dordrecht: Kluwer) p 13
\bibitem{Aharony} Aharony A, Entin-Wohlman O and Imry Y 2003 {\it 
Phys. Rev. Lett.}
{\bf 90} 156802 
\bibitem{BGEW} Berkovits R, Gefen Y and Entin-Wohlman O 1998 {\it Phil. Mag.} B
{\bf 77} 1123
\bibitem{OG97} Oreg Y and  Gefen Y 1997 {\it Phys. Rev.} {\bf B55} 13726 
\bibitem{SOG} Silva A, Oreg Y and  Gefen Y 2002 
{\it Phys. Rev.}  B {\bf 66} 195316
\bibitem{Kim} Kim T-S and Hershfield S 2003
{\it Phys. Rev.} B {\bf 67} 235330 
\bibitem{Leturcq} Leturcq R, Graf D, Ihn T, Ensslin K, Driscoll D D
and Gossard A C 2004 {\it Europhys. Lett.} {\bf 67} 439
\bibitem{Weidenmueller} Weidenm\"{u}ller H A 2002 {\it Phys. Rev.} B 
{\bf 65}, 245322
\bibitem{But99} Taniguchi T and B\"{u}ttiker M 1999 
{\it Phys. Rev.} {\bf B60} 13814
\nonum Levy Yeyati A and B\"{u}ttiker M 2000
{\it Phys. Rev.} B {\bf 62} 7307 
\bibitem{Baltin} Hackenbroich G, Heiss W D and  Weidenm\"{u}ller H A 1997 
{\it Phys. Rev. Lett.} {\bf 79} 127 
\nonum Baltin R, Gefen Y, Hackenbroich G 
and Weidenm\"{u}ller H A 1999 {\it Eur. J. Phys.} B {\bf 10} 119 
\bibitem{Hackenbroich} Hackenbroich G, Heiss W D
and Weidenm\"{u}ller H A 1998 {\it Phil. Mag.} B {\bf 77} 1255
\bibitem{SI} Silvestrov P G and Imry Y 2000 {\it Phys. Rev. Lett.} {\bf 85}
 2565
\bibitem{BvOG} Berkovits R, von Oppen F and Gefen Y 2005 {\it Phys. Rev. Lett.}
 {\bf 94} 076802
\bibitem{Dagotto} B\"{u}sser C A,  Martins G B,  Al-Hassanieh K A, 
Moreo A and Dagotto E 2004 {\it Phys. Rev.} {\bf B70} 245303
\bibitem{Marquardt} Meden V and Marquardt F 2006 {\it Phys. Rev. Lett.} 
{\bf 96} 146801 
\bibitem{Meden} Karrasch G, Hecht T, Oreg Y, von Delft J and  Meden V 
2006 {\it Preprint} cond-mat/0609191
\bibitem{vonDelft} Karrasch G, Hecht T, Weichselbaum A, von Delft J 
Oreg Y and Meden V 2007 {\it New J. Phys.} this issue, p. 
[{\it Preprint} cond-mat/0612490]
\bibitem{Kim06} Kim S and Lee H-W 2006 {\it Phys. Rev.} B {\bf 73} 205319
\bibitem{Lee06} Lee H-W and Kim S 2006 {\it Preprint} cond-mat/0610496
\bibitem{Avi} Kashcheyevs V, Schiller A, Aharony A and Entin-Wohlman O 
2007  {\it Phys. Rev.} B {\bf 75} 115313
\bibitem{ExpInt} Lindemann S, Ihn T, Bieri S, Heinzel T 
Ensslin K, Hackenbroich G, Maranowski K and Gossard A C 
2002 {\it Phys. Rev.} B {\bf 66} 161312
\nonum Johnson A C, Marcus C M, Hanson M P and Gossard A C
2004 {\it Phys. Rev. Lett.} 93 106803  
\nonum Kobayashi K, Aikawa H, Sano A, Katsumoto S and Iye Y
2004 {\it Phys. Rev. } B {\bf 70} 035319 
\bibitem{YG04} K\"{o}nig J and Gefen Y 2005 {\it Phys. Rev.} B {\bf 71} 201308 
\bibitem{Sindel} Sindel M, Silva A,  Oreg Y and von Delft J
2005 {\it Phys. Rev.} B {\bf 72} 125316
\bibitem{Khomskii} Khomskii D I and Kocharyan A N 1976 {\it Solid State
Commun.} {\bf 18} 985  
\nonum Khomskii D I and Kocharyan A N 1976 {\it Zh. Expt. Teor. Fiz.} 
{\bf 71} 767  
[English translation: 1976 {\it Sov. Phys. JETP} {\bf 44} 404]
\bibitem{prb06} Golosov D I and Gefen Y 2006 {\it Phys. Rev.} B {\bf 74} 205316
\bibitem{Oreg} Oreg Y 2007 {\it New J. Phys.}, this issue, p.
\bibitem{Baym} Baym G and Kadanoff L P 1961 {\it Phys. Rev.} {\bf 124} 287 
\bibitem{Goldstein} Goldstein M and Berkovits R 2007  
{\it New J. Phys.} this issue, p. 
[{\it Preprint} cond-mat/0610810] 
\bibitem{berkvoppen} Berkovits R., von Oppen F. and Kantelhardt J. W. 2004
{\it Europhys. Lett.} {\bf 68}, 699 
\end{thebibliography}
\end{document}